\documentclass[twocolumn]{aastex62}
\accepted{ in The Astronomical Journal}
\begin{document}
%\Large
	\title{KELT-11\,b: Abundances of water and constraints on carbon-bearing molecules from the Hubble transmission spectrum}
	
	\correspondingauthor{Q. Changeat}
	\email{quentin.changeat.18@ucl.ac.uk}
	\author[0000-0001-6516-4493]{Q. Changeat}\thanks{These authors contributed equally to this work}
	\affil{Department of Physics and Astronomy \\
		University College London \\
		Gower Street,WC1E 6BT London, United Kingdom}
	\author[0000-0002-5494-3237]{B. Edwards}\thanks{These authors contributed equally to this work}
	\affil{Department of Physics and Astronomy \\
		University College London \\
		Gower Street,WC1E 6BT London, United Kingdom}
	\author[0000-0003-2241-5330]{A.F. Al-Refaie}
	\affil{Department of Physics and Astronomy \\
		University College London \\
		Gower Street,WC1E 6BT London, United Kingdom}
	\author[0000-0001-8587-2112]{M. Morvan}
	\affil{Department of Physics and Astronomy \\
		University College London \\
		Gower Street,WC1E 6BT London, United Kingdom}
	\author[0000-0003-3840-1793]{A. Tsiaras}
	\affil{Department of Physics and Astronomy \\
		University College London \\
		Gower Street,WC1E 6BT London, United Kingdom}
	\author[0000-0002-4205-5267]{I.P. Waldmann}
	\affil{Department of Physics and Astronomy \\
		University College London \\
		Gower Street,WC1E 6BT London, United Kingdom}
	\author[0000-0001-6058-6654]{G. Tinetti}
	\affil{Department of Physics and Astronomy \\
		University College London \\
		Gower Street,WC1E 6BT London, United Kingdom}
	
%\maketitle

%%%%%%%%%%%%%%%%%%%%%%%%%%%%%%%%%%%%%%%%%%%%%%%%%%%%%%%%%%%%%%%%%%%%%%%%%%%%%%%%
\begin{abstract}

   In the past decade, the analysis of exoplanet atmospheric spectra has revealed the presence of water vapour in almost all the planets observed, with the exception of a fraction of overcast planets \citep{Tsiaras_pop_study_trans, Fu_2017_stat, Barstow_10HJ, sing_exoplanets, Pinhas_ten_HJ_clouds}. Indeed, water vapour presents a large absorption signature in the wavelength coverage of the Hubble Space Telescope's (HST) Wide Field Camera 3 (WFC3), which is the main space-based observatory for  atmospheric studies of exoplanets, making its detection very robust. However, while carbon-bearing species such as methane, carbon monoxide and carbon dioxide are also predicted from current chemical models \citep{2015_venot_diseq}, their direct detection and abundance characterisation has remained a challenge. Here we analyse the transmission spectrum of the puffy, clear hot-Jupiter KELT-11\,b from the HST WFC3 camera. We find that the spectrum is consistent with the presence of water vapor and an additional absorption at longer wavelengths than 1.5$\mu$m, which could well be explained by a mix of carbon bearing molecules. CO$_2$, when included is systematically detected. One of the main difficulties to constrain the abundance of those molecules is their weak signatures across the HST WFC3 wavelength coverage, particularly when compared to those of water. Through a comprehensive retrieval analysis, we attempt to explain the main degeneracies present in this dataset and explore some of the recurrent challenges that are occurring in retrieval studies (e.g: the impact of model selection, the use of free vs self-consistent chemistry and the combination of instrument observations).   %While the extrapolation of C/O ratio and metallicity indicators is subject to strong biases, the recovered abundances may indicate a carbon enriched atmosphere. The data was analysed using the publicly available software Iraclis and TauREx. 
   Our results make this planet an exceptional example of chemical laboratory where to test current physical and chemical models of hot-Jupiters' atmospheres.\\

\end{abstract}
%%%%%%%%%%%%%%%%%%%%%%%%%%%%%%%%%%%%%%%%%%%%%%%%%%%%%%%%%%%%%%%%%%%%%%%%%%%%%%%%

\section{Introduction}

Transmission and emission spectroscopy have formed the cornerstone of exoplanet atmospheric characterisation, enabling the discovery of water in many planets \citep{Tinetti_water, crouzet_hd189,Tsiaras_pop_study_trans}. While the detection of water is now routine for the hot-Jupiter class of planets, other molecules such as the carbon species remains challenging with current space instrumentation. With a few exceptions \citep{Swain_2008,Swain_2009}, most claims for the carbon species from space have been based on additional absorption in the infrared Spitzer photometric bands  \citep{Madhusudhan_2010_W12,stevenson_w43_2,  2016_line_hd209, Gandhi_2019_hd209} or from ground based observations using either direct imaging \citep{Macintosh_2015,Barman_2015,Lacour_2019} or high-dispersion techniques \citep{ Snellen_2010_HR_HD209, de_Kok_2013, Konopacky_2013_coHR, Brogi_2017_combinedRet}. 
While very valuable, these detections often lack a reference baseline or require the combination of multiple instruments, each with different systematics, which may limit the determination of absolute abundances or may lower the detection significance \citep{Yip_lightcurve, Brogi_2019_ret}.

To analyse those exoplanet spectra, inverse retrieval techniques are often used \citep{rodger_retrievals,irwin2008, Madhu_retrieval_method,chimera, Waldmann_taurex1,Waldmann_taurex2,Gandhi_2019_hd209,Mollire_petitrad, al-refaie_taurex3, Zhang_platon, min2020arcis}. These techniques explore the information content in an exoplanet spectrum and map the parameters space of possible solutions. In the last few years, it has become a standard practice to perform atmospheric retrievals using `self-consistent' chemical models as opposed to `free' chemical models. In free retrievals, the chemical composition of the atmosphere is retrieved using parametric profiles, which are not assuming prior knowledge. For example profiles can be assumed constant with altitude or use more complex parametric descriptions when required \citep{Changeat_2019_2l,Parmentier_2018}. Self-consistent models rely on simplifying assumptions (atmosphere in thermo-chemical equilibrium) to reduce the number of free parameters in the model and provide a more complex chemical structure (variation of chemistry with altitude). For exoplanets, it remains a strong assumption of the physical and chemical state of the atmosphere that can lead to strong biases \citep{2015_venot_diseq, Changeat_2019_2l, Changeat_2020_alfnoor, anisman2020wasp117}.

The planet KELT-11\,b was discovered in 2016 orbiting a bright G star (Kmag = 6.122), with an orbital period of 4.736 days \citep{Pepper_2017_kelt11}. Due to its very low density (0.093 g.cm$^{-3}$), it was immediately associated with a very large scale height and was predicted to become one of the ``benchmark'' planets for atmospheric characterisation. Further observations from the ground and the Spitzer Space Telescope refined the orbital and star parameters \citep{Beatty_2017_kelt11}. A recent paper \citep{Zak_kelt11_2019} analysed the high resolution data from HARPS  in the search for sodium, hydrogen and lithium. They saw no evidence for these species, and they attributed the non-detection to the possible presence of high altitude clouds. They  also reported a low stellar activity of the host star, a result obtained by monitoring the Ca I and Mg I lines.

Here we present the analysis of a single transit of KELT-11\,b from the Hubble Space Telescope. We first describe the way our analysis was carried using the public pipeline Iraclis and the public retrieval code TauREx3. Then we show the results from our free exploration of the atmospheric properties, showing that multiple solutions can reflect the information content in this spectrum. We then discuss the use of self-consistent chemical models in atmospheric retrievals and the combination of observations using the complementary TESS data. Our results are also compared with a recent independent analysis of the same dataset from \cite{coln2020unusual}.

\section{Methodology}

\subsection{Extraction of Planetary Spectrum}

%The HST data was taken with the G141 grism (1.088-1.68 $\mu$m) of the Wide Field Camera 3 (WFC3) in April 2018 (PN: 15225, PI: Knicole Colon). We obtained the publicly available data from the HST MAST archive\footnote{website: https://archive.stsci.edu/hst/search.php}. We used our publicly available tools to perform the end-to-end analysis from the raw data to the atmospheric parameters. The HST data was reduced, and the light curves fitted, using the Iraclis software \citep{Tsiaras_2016_iraclis}. We then used our Bayesian retrieval code TauREx3 \citep{al-refaie_taurex3} to extract and analyse the molecular content of this atmosphere.

%A single transit of KELT-11\,b was taken for the HST proposal 15255 (principal investigator: Knicole Colon), and the data are available through the Mikulski Archive for Space Telescopes (MAST). The visit consisted of 9 HST orbits with the G141 infrared grism of the WFC3 camera (1.088-1.68 $\mu$m), in the spatial scanning mode. During an exposure using the spatial scanning mode, the instrument slews along the cross-dispersion direction, allowing for longer exposure times and increased signal-to-noise ratio, without the risk of saturation \citep{deming_hd209}. Both forward (increasing row number) and reverse (decreasing row number) scanning were used for these observations to increase the duty cycle.

A single transit of KELT-11\,b was taken with the G141 grism (1.088-1.68 $\mu$m) of the Wide Field Camera 3 (WFC3) in April 2018 (PN: 15225, PI: Knicole Colon). We obtained the publicly available data from the HST MAST archive\footnote{website: https://archive.stsci.edu/hst/search.php}. We used our publicly available tools to perform the end-to-end analysis from the raw data to the atmospheric parameters. The HST data was reduced, and the light curves fitted, using the Iraclis software \citep{Tsiaras_2016_iraclis}. We then used our Bayesian retrieval code TauREx3 \citep{al-refaie_taurex3} to extract and analyse the molecular content of this atmosphere.

The visit consisted of 9 HST orbits with the G141 infrared grism of the WFC3 camera (1.088-1.68 $\mu$m), in the spatial scanning mode. During an exposure using the spatial scanning mode, the instrument slews along the cross-dispersion direction, allowing for longer exposure times and increased signal-to-noise ratio, without the risk of saturation \citep{deming_hd209}. Both forward (increasing row number) and reverse (decreasing row number) scanning were used for these observations to increase the duty cycle.

The detector settings were SUBTYPE = SQ512SUB, SAMP$_{SEQ}$ = SPARS25, NSAMP = 4, APERTURE = GRISM512, and the scanning speed was 0.96 "$s^{-1}$. The final images had a total exposure time of 46.695518 s, a maximum signal level of 36,000 electrons per pixel and a total scanning length of 51.312 ". For calibration purposes, a 2.559081 s non-dispersed (direct) image of the target was taken at the beginning of each orbit, using the F130N filter and the following settings: SUBTYPE = SQ512SUB, SAMP$_{SEQ}$ = RAPID, NSAMP = 4, APERTURE = GRISM512.

We carried out the analysis of the transit using Iraclis, our highly-specialised software for processing WFC3 spatially scanned spectroscopic images \citep{Tsiaras_2016_iraclis, Tsiaras_2016_55cnc, Tsiaras_pop_study_trans,Tsiaras_2019_k2-18}. The reduction process included the following steps: zero-read subtraction, reference-pixels correction, nonlinearity correction, dark current subtraction, gain conversion, sky background subtraction, calibration, flat-field correction and bad-pixels/cosmic-rays correction. Then we extracted the white (1.088-1.68 $\mu$m) and the spectral light curves from the reduced images, taking into account the geometric distortions caused by the tilted detector of the WFC3 infrared channel.

We fitted the light curves using our transit model package PyLightcurve, with the transit parameters from \cite{Beatty_2017_kelt11} and limb-darkening coefficients calculated based on the PHOENIX \citep{phoenix} model (see tables in Appendix 1), the nonlinear formula and the stellar parameters, also from \cite{Beatty_2017_kelt11}. These were computed using ExoTETHyS \cite{morello_exotethys}. During our fitting of the white light curve, the planet-to-star radius ratio and the mid-transit time were the only free parameters, along with a model for the systematics \citep{Kreidberg_GJ1214b_clouds, Tsiaras_2016_iraclis}. 

It is common for WFC3 exoplanet observations to be affected by two kinds of time-dependent systematics: the long-term and short-term `ramps' \citep[e.g.][]{Kreidberg_wasp12, Evans_wasp121, 2016_line_hd209}. The first affects each HST visit and usually has a linear behaviour, while the second affects each HST orbit and is modelled as having an exponential behaviour. The formula we used for the white light curve systematics (Rw) was the following:

\begin{equation}
    R_w(t) = n^{scan}_w(1-r_a(t - T_0))(1-r_{b1}e^{-r_{b2}(t-t_)})
\end{equation}

\noindent where $t$ is time, $n^{scan}_w$ is a normalisation factor, $T_0$ is the mid-transit time, $t_o$ is the time when each HST orbit starts, $r_a$ is the slope of a linear systematic trend along each HST visit and ($r_{b1},r_{b2}$) are the coefficients of an exponential systematic trend along each HST orbit. The normalization factor we used ($n^{scan}_w$) was changed to $n^{for}_w$ for upward scanning directions (forward scanning) and to $n^{rev}_w$) for downward scanning directions (reverse scanning). The reason for using different normalization factors is the slightly different effective exposure time due to the known upstream/downstream effect \citep{McCullough_wfc3}. 

\begin{figure}
    \includegraphics[width = \columnwidth]{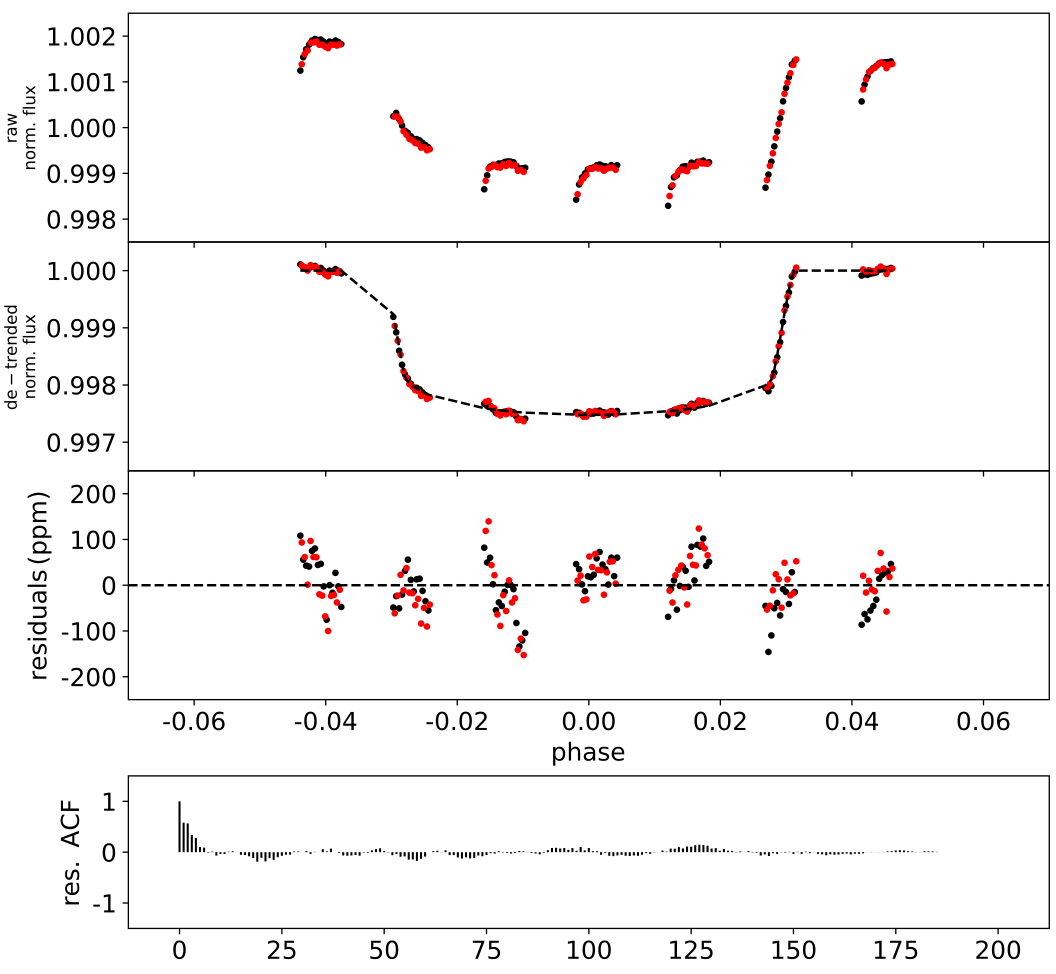}
    \caption{White light curve fit for the transit of KELT-11\,b. First panel: raw light curve, after normalisation. Second panel: light curve, divided by the best fit model for the systematics. Third panel: residuals for best-fit model. Fourth panel: auto-correlation function of the residuals. Black data points are for forwards scans while data from reverse scans is indicated in red.}
    \label{fig:white}
\end{figure}

We fitted the white light curves using the formulae above and the uncertainties per pixel, as propagated through the data reduction process. However, it is common in HST/WFC3 data to have additional scatter that cannot be explained by the ramp model. For this reason, we scaled up the uncertainties in the individual data points, for their median to match the standard deviation of the residuals, and repeated the fitting \citep{Tsiaras_pop_study_trans}. We found the orbital parameters literature orbital parameters \citep{Beatty_2017_kelt11} to provide an excellent fit to to the data and thus the only free parameters in our white fitting, other than the HST systematics, were the mid transit time and the planet-to-star radius ratio. We show the white light curve fitting resulting from our spectrum extraction step in Figure \ref{fig:white}. 

In our analysis, we found that the measured mid-transit time had drifted from the expected ephemeris. We therefore used this observation, along with data from the Transiting Exoplanet Satellite (TESS, \citet{ricker_tess}), to refine the ephemeris of KELT-11\,b.

Next, we fitted the spectral light curves with a transit model (with the planet-to-star radius ratio being the only free parameter) along with a model for the systematics ($R_\lambda$) that included the white light curve (divide-white method \citep{Kreidberg_GJ1214b_clouds}) and a wavelength-dependent, visit-long slope \citep{Tsiaras_2016_iraclis}. 

\begin{equation}
    R_\lambda(t) = n^{scan}_\lambda(1-\chi_\lambda(t-T_0))\frac{LC_w}{M_w}
\end{equation}{}

\noindent where $\chi_\lambda$ is the slope of a wavelength-dependent linear systematic trend along each HST visit, $LC_w$ is the white light curve and $M_w$ is the best-fit model for the white light curve. Again, the normalization factor we used ($n^{scan}_\lambda$) was changed to ($n^{for}_\lambda$) for upward scanning directions (forward scanning) and to ($n^{for}_\lambda$) for downward scanning directions (reverse scanning). Also, in the same way as for the white light curves, we performed an initial fit using the pipeline uncertainties and then refitted while scaling these uncertainties up, for their median to match the standard deviation of the residuals. The final extracted spectrum is given in Appendix 2.

\begin{figure}
    \centering
    \includegraphics[width = \columnwidth]{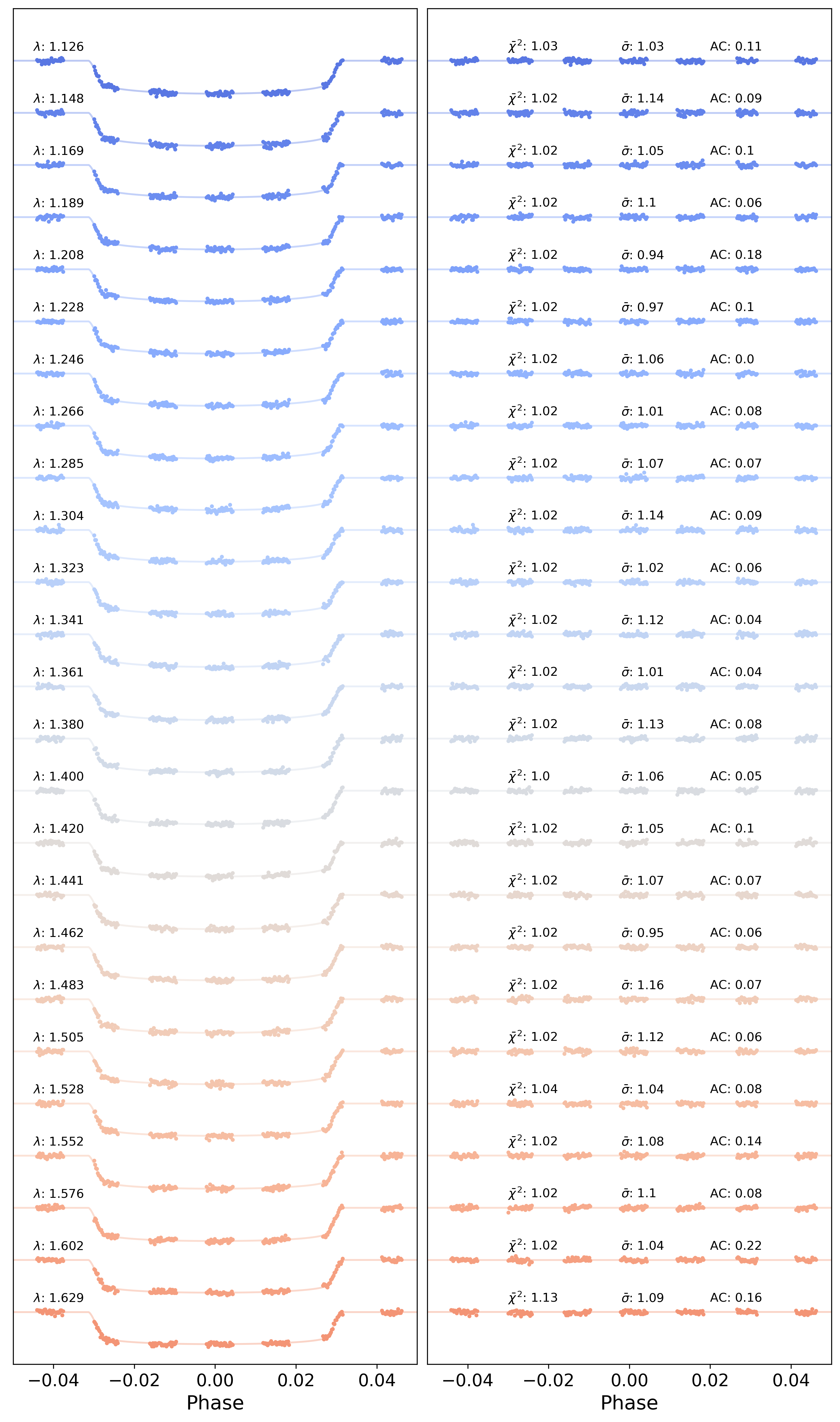}
    \caption{Spectral light curves fitted with Iraclis for the transmission  spectra where, for clarity, an offset has been applied. Left: the detrended spectral light curves with best-fit model plotted. Right: residuals from the fitting with values for the Chi-squared ($\chi^2$), the standard deviation of the residuals with respect to the photon noise ($\bar{\sigma}$) and the auto-correlation (AC).}
    \label{fig:spec_lc}
\end{figure}

\subsection{Spitzer Transit Observation}

A transit observation was taken with Spitzer's Infrared Camrea Array (IRAC) at at $3.6$ $\mu m$ (PID: 12096, principal investigator: T. Beatty). We decided not to include it in this present analysis because, as stated in \cite{Beatty_2017_kelt11}, the transit occurred earlier than expected, meaning the pre-ingress part of the light curve is missing. As the `ramp' effect is especially pronounced during the settling of IRAC observations (see e.g. \citealp{agol_2010_ramp}), the preprocessing of this light curve by any of the standard detrending techniques \citep[e.g.][]{mariooo} would likely lead to larger uncertainties in the retrieved transit depth\footnote{In fact, the initial exposures of exoplanet light curves with Spitzer, where the telescope is settling, are very often discarded on account of the steepest ramps on these portions}. This would in turn reduce the chances of a robust combination between observations from the different instruments \citep{Yip_lightcurve, giovanni2020}.

\subsection{TESS Data Reduction \& Ephemeris Refinement}

\begin{figure}
    \centering
    \includegraphics[width=0.9\columnwidth]{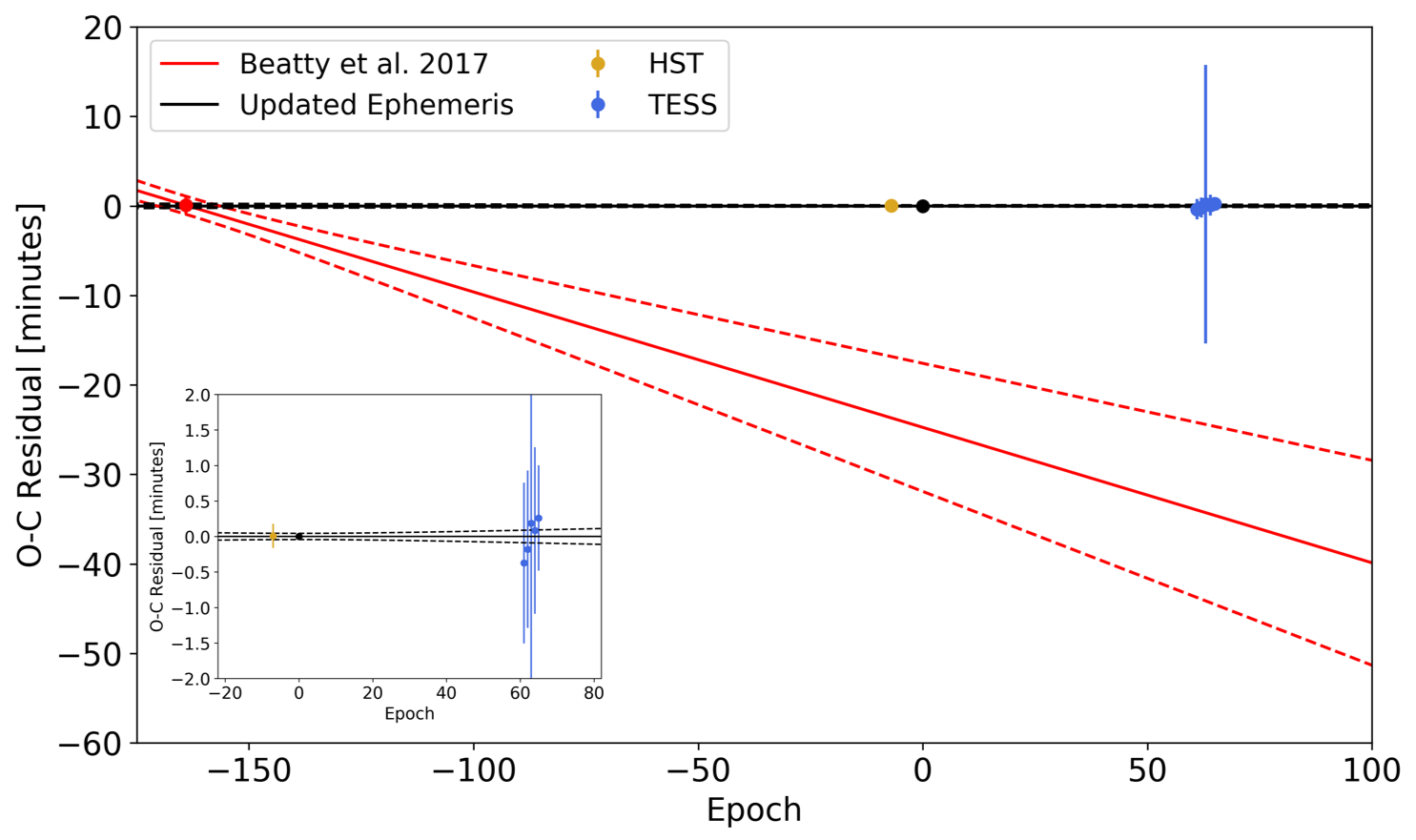}
    \caption{O-C residuals for KELT-11\,b. The literature ephemeris \cite{Beatty_2017_kelt11} are shown in red, the HST transit in gold and the TESS data in blue. The black line denotes the new ephemeris of this work with the dashed lines showing the associated 1$\sigma$ uncertainties. The inset shows a zoomed plot to highlight the accuracy of the transit times measured here. We note that the third TESS light curve had a high uncertainty on the transit mid time due to an interruption in the observation which caused egress to be missed.}
    \label{ephm_refine}
\end{figure}

Accurate knowledge of exoplanet transit times is crucial for atmospheric studies. To ensure that KELT-11\,b can be observed in the future, we used the HST white light curve mid time, along with data from TESS, to update the ephemeris of the planet. TESS data is publicly available through the MAST archive and we use the pipeline from \cite{edwards_orbyts} to download, clean and fit the 2 minute cadence Pre-search Data Conditioning (PDC) light curves \citep{smith_pdc,stumpe_pdc1,stumpe_pdc2}. KELT-11\,b was studied in Sector 9 and, after excluding bad data, we recovered 5 transits. These were fitted individually with the planet-to-star radius ratio $R_p/R_s$, reduced semi-major axis ($a/R_s$), inclination ($i$) and transit mid time ($T_0$) as free parameters. The observed minus calculated (O-C) residuals are shown in Figure \ref{ephm_refine} along with the detrended data, the best-fit model and the residuals for each transit. We calculated the ephemeris to be P = 4.73620495$\pm$0.00000086 days and T$_0$ = 2458260.168608$\pm$0.000030 BJD$_{TDB}$ where P is the period, T$_0$ is the mid-time of the transit and BJD$_{TDB}$ is the barycentric Julian date in the barycentric dynamical time standard. The mid times used and the updated parameters are given in Table \ref{tab:mid_times}.

\begin{table*}
    \centering
    \begin{tabular}{ccc}\hline\hline
    Epoch & Transit Mid Time [$BJD_{TDB}$] & Reference\\\hline\hline
    -164 & 2457483.431 $\pm$ 0.0007 & \cite{Beatty_2017_kelt11}\\
    -7 & 2458227.015148 $\pm$ 0.000111 & This Work \\
    61 & 2458549.077357 $\pm$ 0.000524 & This Work \\
    62 & 2458553.813428 $\pm$ 0.000616 & This Work \\
    63 & 2458558.54938 $\pm$ 0.006887 & This Work \\
    64 & 2458563.285654 $\pm$ 0.000481 & This Work \\
    65 & 2458568.021737 $\pm$ 0.000486 & This Work \\ \hline\hline
    \multicolumn{3}{c}{Derived Values}\\\hline\hline
    Period (P) & 4.73620495$\pm$0.00000086 & days \\
    Transit Mid Time (T$_{0}$) & 2458260.168608$\pm$0.000030 & BJD$_{TDB}$ \\\hline\hline
    \end{tabular}
    \caption{Transit mid times used to refine the ephemeris of KELT-11 b in this work.}
    \label{tab:mid_times}
\end{table*}{}

\subsection{Retrieval setup}

The observed spectrum (Appendix 2) was analysed using our Bayesian retrieval framework TauREx\,3 \citep{Waldmann_taurex1,Waldmann_taurex2,al-refaie_taurex3,al-refaie_taurex3.1}, which was recently benchmarked \citep{Barstow_2020_compa} against the other retrieval codes NEMESIS \citep{irwin2008} and CHIMERA \citep{chimera}. We utilised the absorption cross-sections from the ExoMol database \citep{tennyson2012exomol, Tennyson_exomol, tennyson2020_2020, chubb_2020_exomolop} and explored the parameter space with the algorithm MultiNest \citep{Feroz_multinest} {with 750 live points and an evidence tolerance of 0.5}. We adopted uniform priors for all the free parameters.

Our atmosphere is a 1-dimensional model consisting of 100 layers, covering the pressures from 10 bar to 10$^{-9}$ bar, equally spaced in log-scale. For the trace gases we considered the molecules: H$_2$O \citep{barton_h2o, polyansky_h2o}, CH$_4$ \citep{hill_xsec, exomol_ch4}, CO \citep{li_co_2015}, CO$_2$ \citep{rothman_hitremp_2010}, C$_2$H$_2$ \citep{2016_WILZEWSKI_C2H2}, C$_2$H$_4$ \citep{2018_Mant_C2H4}, HCN \citep{Barber_2013,Harris_2006}, TiO \citep{McKemmish_TiO_new} , VO \citep{McKemmish_2016_vo} and FeH \citep{Bernath_2020_FeH}. 

In order to avoid biases from chemical assumptions, we considered free chemistry for the main result section. For completeness and due to the particularly high signal to noise ratio of the KELT-11\,b spectrum, we still discuss the case of equilibrium chemistry in the discussion section. In free chemistry, we fit each of the molecule abundances in volume mixing ratios with log-prior bounds from -12 to -1. The rest of the atmosphere is composed of H$_2$ and He for which the ratio is fixed to solar values. On top of the molecular absorption, we include opacity from Rayleigh scattering \citep{cox_allen_rayleigh} and Collision Induced Absorption (CIA) processes from H$_2$-H$_2$ \citep{abel_h2-h2} and H$_2$-He \citep{abel_h2-he} pairs. The planetary mass was fixed to the literature values in all the performed retrievals, since it is poorly constrained by HST observations \citep{Changeat_2020_mass}. 

In order to model clouds we included a Grey opacity - a fully opaque atmosphere above a given pressure -  and attempted to recover the top pressure of this cloud deck P$_{c}$. Finally, we fit for the planet radius R$_p$ at 10 bar with bounds 0.9 R$_J$ - 1.6 R$_J$ and an isothermal temperature profile $T$ with bounds 500K - 2500K. In transit, the temperature affects mostly the atmospheric scale height and the narrow wavelength range of HST does not allow to recover the thermal structure precisely \citep{Rocchetto_biais_JWST, Changeat_2019_2l}.

In the results section, we investigate the information that can be extracted from this spectrum by running several free retrieval models: \\

- A `base' retrieval composed of water and the main carbon based molecules (CH$_4$, CO and CO$_2$). It is a conservative model as the considered molecules are expected to be present in the atmosphere of KELT-11\,b (Section 3.1). \\

- An `extended' retrieval model. Since the observed spectrum contains additional absorption in the longer wavelengths, we investigate a larger range of carbon compounds. HCN, C$_2$H$_2$ and C$_2$H$_4$ are added to the 'base' setup (Section 3.2). \\

- A `water only' retrieval, which only contains absorption from water vapor (Section 3.3). This allows us to statistically assess the relevance of the carbon bearing species detections in the `base' and `extended' models.  \\

- A `full' retrieval scenario, which also includes the near-optical absorbers TiO, VO and FeH. This addition was motivated from studying the combined TESS+HST spectrum (Section 3.4). \\

For model comparison, we provide the relative Global log evidence (log E) of each solution in Table \ref{tab:evidence_compa}. These are relative to a standard flat line model (log E$_{flat}$ = 146.6) built by removing all wavelength dependant absorption and fitting only the radius, temperature and clouds \citep{Tsiaras_pop_study_trans}. In the discussion section, we provide complementary retrievals with the aim to use KELT-11\,b as an example to illustrate particular aspects of retrieval study: \\

- The `ACE' retrieval uses an equilibrium chemistry scheme from \cite{Agundez_2dchemical_HD209_HD189}. For this run, the only two chemical free parameters are the metallicity (log M) and C/O ratio (Section 4.1). \\

- A `combined' retrieval is also performed. This uses the `full' setup on a spectrum combining the HST spectrum with the TESS photometric point (Section 4.2). \\

\section{Results}

\begin{figure*}
    \centering
    \includegraphics[width=0.8\textwidth]{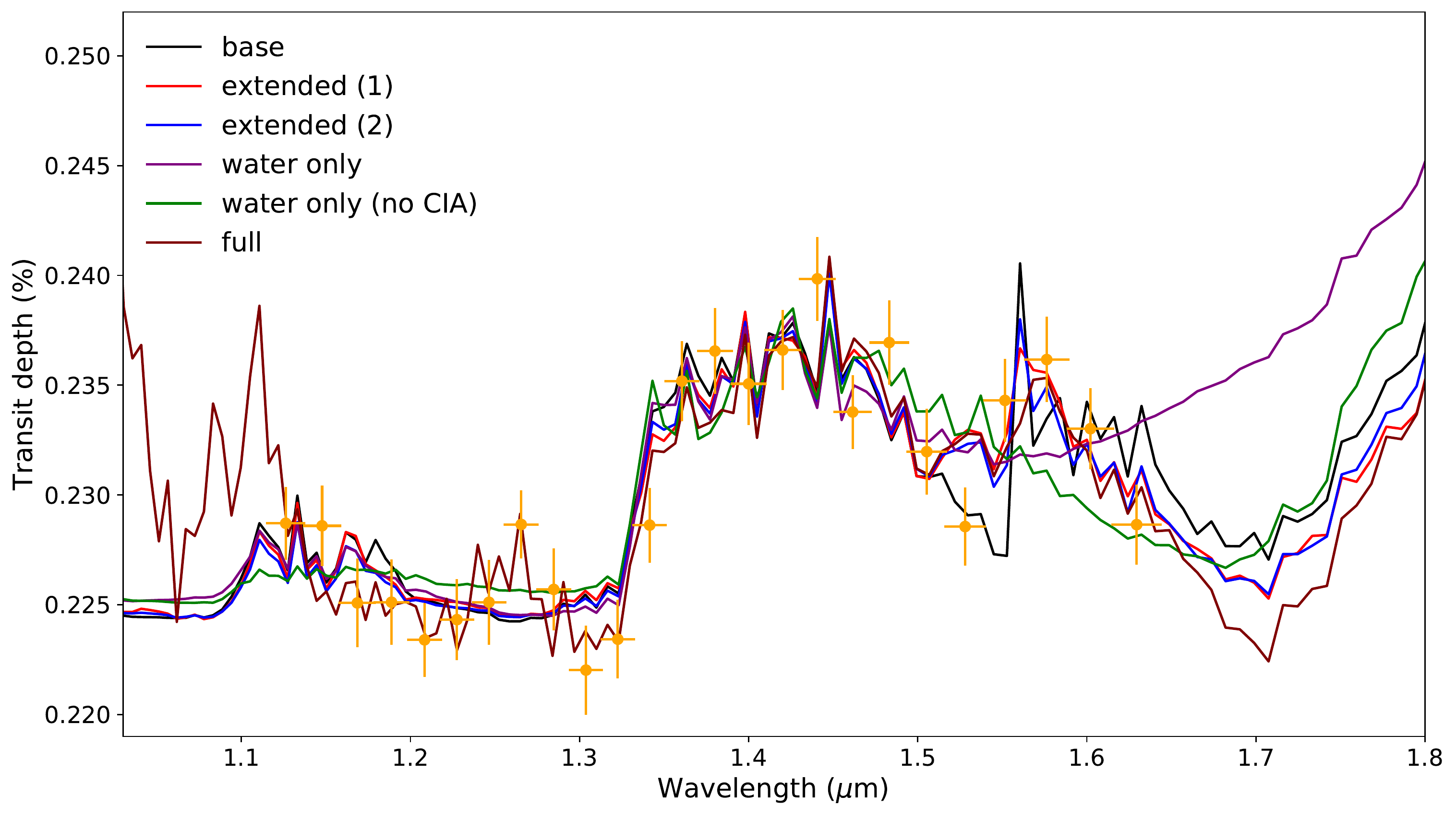}
    \caption{Comparison of the best-fit spectra from the different retrievals on the KELT-11\,b reduced HST data (orange). The feature seen after 1.5$\mu$m is better reproduced by models including carbon species but all the models provide a decent fit to this spectrum.}
    \label{fig:spectrum_compa}
\end{figure*}

From our retrieval exploration, we find that the free models provide a reasonable fit to the observed spectrum. Figure \ref{fig:spectrum_compa} compares the best fit spectra for all the considered retrievals. Similarly, Table \ref{tab:evidence_compa} provides a summary of the retrieved parameters for each model as well as the global log evidence relative to the flat line model. This indicator describes the significance to which an atmospheric signal is detected and allows us to compare the tested models \citep{Kass1995bayes, Tsiaras_pop_study_trans}. 

The selection of the best model, amongst the ones tested, is difficult since their respective log E lie within the same range (variations of less than 3). The `full' model provides the best statistical fit but is only marginally better than the other solutions. In all the tested scenarios, the temperature seems consistent (except in the `ACE'  retrieval) around 1300K. We note that all those solutions agree on the presence of water vapour in the atmosphere of KELT-11\,b and that some combination of additional species is necessary to fully explain the peculiar absorption after 1.5$\mu$m. The recovered abundance of water in our models is sub-solar, as also suggested by a recent independent study from \cite{coln2020unusual}. When included in the retrieval, CO$_2$ is systematically recovered, despite abundances that variate depending on the model considered. For the other species (CO and HCN), their detection depends on the model considered (see the next sections for the details). Here, we detail the 4 types of retrievals we attempted for this planet.

\begin{table*}
    \scriptsize
    \centering
    \begin{tabular}{ccccccc}
    \hline\hline
    Parameter & Base & Extended (1) & Extended (2) & Water only & Full & ACE \\\hline\hline
    Rp & 1.19$^{+0.01}_{-0.03}$ [1.18] & 1.13$^{+0.03}_{-0.04}$ [1.13] & 1.18$^{+0.01}_{-0.03}$ [1.14] & 1.18$^{+0.01}_{-0.01}$ [1.18] & 1.10$^{+0.03}_{-0.05}$ [1.09] & 0.94$^{+0.03}_{-0.03}$ [1.01] \\
    T & 1334$^{+206}_{-167}$ [1216]& 1194$^{+188}_{-155}$ [1169] & 1273$^{+196}_{-150}$ [1349] & 1423$^{+120}_{-83}$ [1504] & 1307$^{+172}_{-141}$ [1347] & 2462$^{+76}_{-125}$ [2398] \\ 
    Pc & 5.0$^{+0.7}_{-0.8}$ [4.2] & 2.7$^{+0.6}_{-0.6}$ [2.7] & 4.4$^{+1.0}_{-0.6}$ [3.6] & 5.2$^{+0.5}_{-0.5}$ [4.7] & 2.5$^{+0.6}_{-0.9}$ [2.3] & 1.8$^{+0.4}_{-0.4}$ [2.9] \\
    log H$_2$O & -5.9$^{+0.4}_{-0.2}$ [-5.3] & -4.0$^{+0.7}_{-0.7}$ [-3.9]  & -5.7$^{+0.5}_{-0.3}$ [-5.1] & -6.2$^{+0.1}_{-0.1}$ [-6.2] &  -3.6$^{+0.6}_{-0.7}$ [-3.3] & - \\
    log CO & -3.8$^{+1.2}_{-4.3}$ [-2.0] & -6.9$^{+3.4}_{-3.3}$ [-2.3] & -4.5$^{+1.6}_{-4.7}$ [-2.9] & - &  -7.3$^{+3.2}_{-2.9}$ [-11.6] & - \\
    log CH$_4$ & -9.7$^{+1.7}_{-1.6}$ [-7.9] & -8.4$^{+2.2}_{-2.3}$ [-6.4] & -9.5$^{+1.7}_{-1.6}$ [-11.4] & - &  -8.1$^{+2.4}_{-2.4}$ [-10.1] & - \\
    log CO$_2$ & -6.5$^{+1.8}_{-3.7}$ [-4.4] & -3.0$^{+0.7}_{-0.9}$ [-2.7] & -4.9$^{+0.8}_{-4.0}$ [-4.0] & - & -3.0$^{+0.6}_{-2.7}$ [-2.1] & - \\
    log HCN & - & -4.1$^{+0.7}_{-0.7}$ [-4.1] & -6.5$^{+1.1}_{-3.4}$ [-5.4] & - & -3.7$^{+0.7}_{-0.8}$ [-3.5] & - \\
    log C$_2$H$_2$ & - & -9.2$^{+1.9}_{-1.8}$ [-8.6] & -9.6$^{+1.5}_{-1.5}$ [-10.4] & - &  -8.8$^{+2.1}_{-2.0}$ [-8.7] & - \\
    log C$_2$H$_4$ & - & -8.6$^{+2.2}_{-2.1}$ [-9.5] & -9.5$^{+1.6}_{-1.6}$ [-11.8] & - &  -8.4$^{+2.6}_{-2.3}$ [-9.5] & - \\
    log TiO & - & - & - & - & -5.1$^{+0.6}_{-0.8}$ [-4.9] & - \\
    log VO & - & - & - & - & -10.0$^{+1.3}_{-1.2}$ [-9.6] & - \\
    log FeH & - & - & - & - & -8.8$^{+1.7}_{-2.0}$ [-6.6] & - \\
    \hline\hline
    log M & -0.7 [1.1] & 0.4 [1.0] & -1.2 [0.2] & -3.3 [-3.3] &  0.4 [1.29] & 0.3$^{+0.5}_{-0.7}$ [0.3] \\
    C/O & 0.99 [1.00] & 0.51 [0.78] & 0.76 [0.93] & - &  0.53 [0.50] & 0.77$^{+0.1}_{-0.3}$ [0.88] \\
    \hline\hline
    $\Delta$ log E & 65.4 & 66.5 & 64.9 & 64.5 & 66.9 & 59.3  \\\hline\hline
    \end{tabular}
\caption{Summary of our different retrieval scenarios. For each parameter we provide the median and 1$\sigma$ retrieved parameters as well as the value from the maximum-a-posteriori (MAP) in bracket. A large difference between the median and the MAP highlights a parameter that did not converge or a non-Gaussian behaviour. We also derive the metallicity (log M) and carbon to oxygen ratio (C/O) following \citep{MacDonald_2019}. The last line provides the global log evidence of each retrieval relative to a flat line model. This is built using $\Delta$ log E = log E - log E$_{flat}$.}
\label{tab:evidence_compa}
\end{table*}

\subsection{Base retrieval results}

\begin{figure}
    \centering
    \includegraphics[width=\columnwidth]{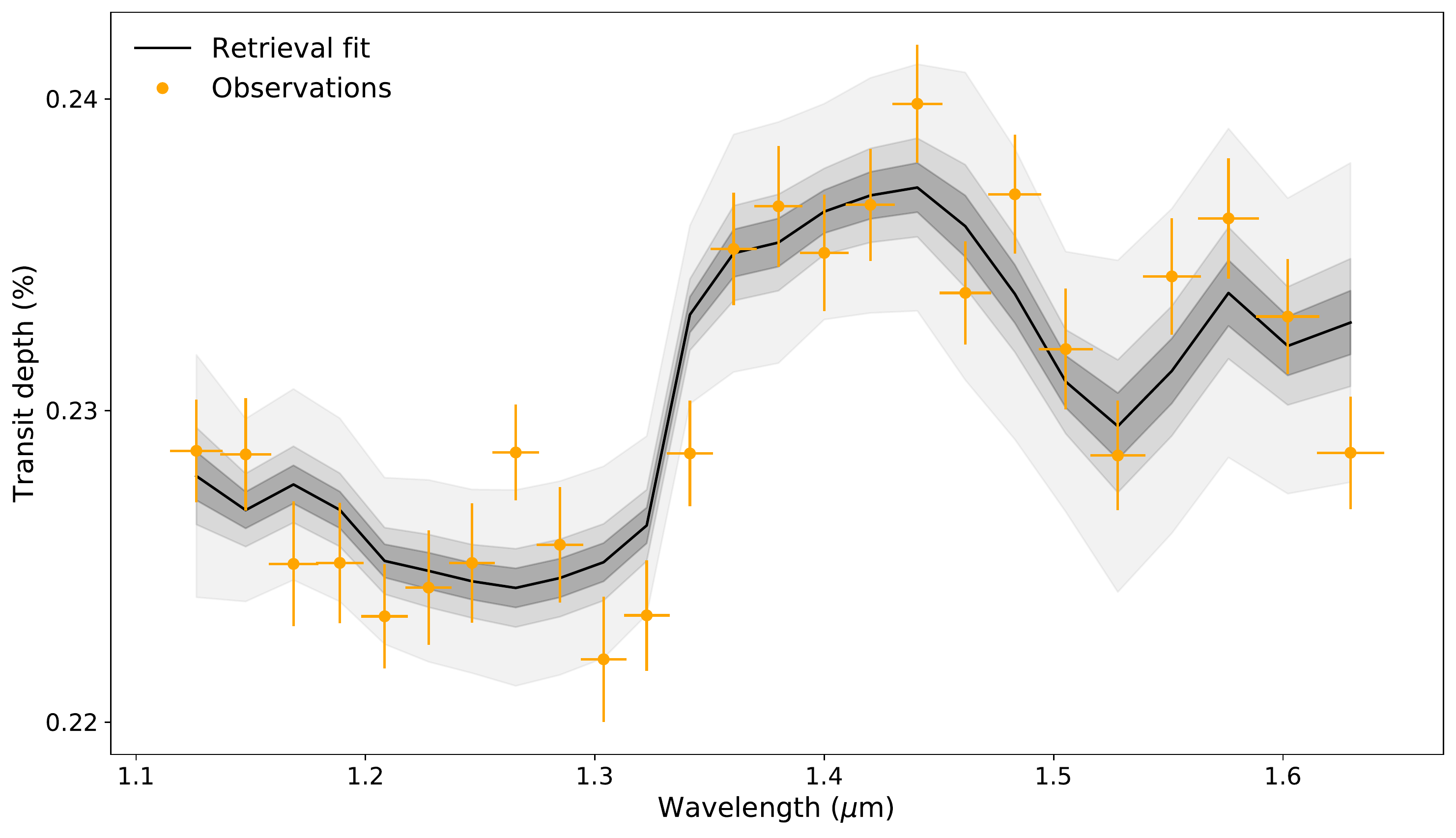}
    \includegraphics[width=\columnwidth]{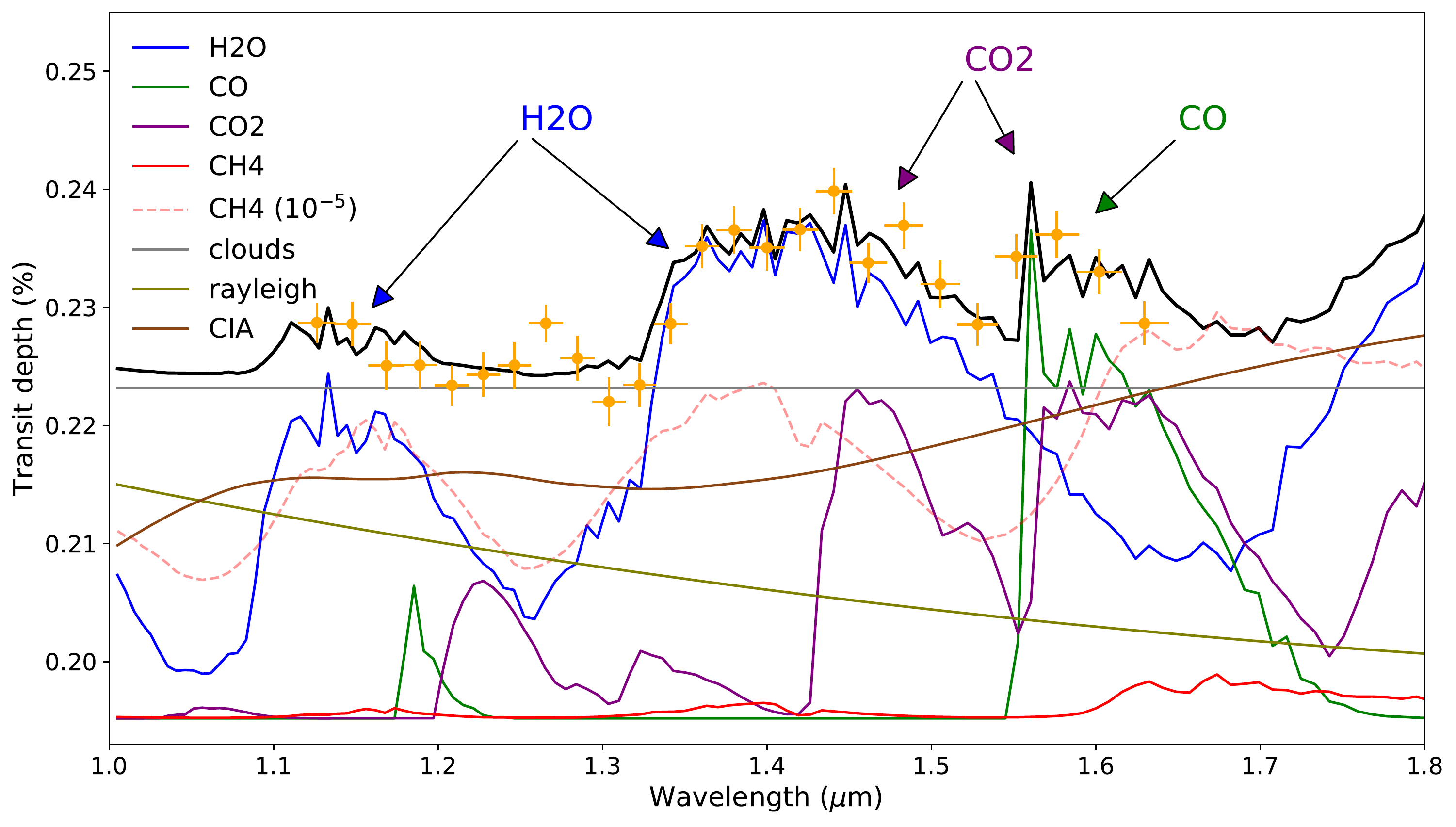}
    \caption{Extracted spectrum of KELT-11\,b observed by HST with 1$\sigma$ error bars (yellow) and fitting results from our base retrieval analysis. Top: Best-fit spectrum from our retrieval analysis (black) with the 1$\sigma$, 2$\sigma$ and 5$\sigma$ regions (shaded dark). Bottom: Best-fit contributions from the different absorbing species. Since CH$_4$ does not contribute in the best-fit model, we show the contribution it would have for an abundance of 10$^{-5}$ with the dashed red line.}
    \label{fig:spectrum}
\end{figure}

Our first analysis of KELT-11\,b uncovers the presence of large molecular signatures in the atmosphere, as seen in the spectral modulations in Figure \ref{fig:spectrum}.

%\begin{figure*}
%    \centering
%    \includegraphics[width = 0.8\textwidth]{post2.pdf}
%    \caption{Correlation maps for the free parameters in our Bayesian retrieval analysis. All the parameters are well constrained. The posterior distributions indicate the detection and precise abundances for H$_2$O, CO and CO$_2$. They also show tight constraints on the radius and the temperature of KELT-11\,b. Methane is not detected and only a upper limit on this abundance can be deduced. We do not recover evidence for the presence of clouds.}
%    \label{fig:post}
%\end{figure*}

\begin{figure*}
    \centering
    \includegraphics[width=\textwidth]{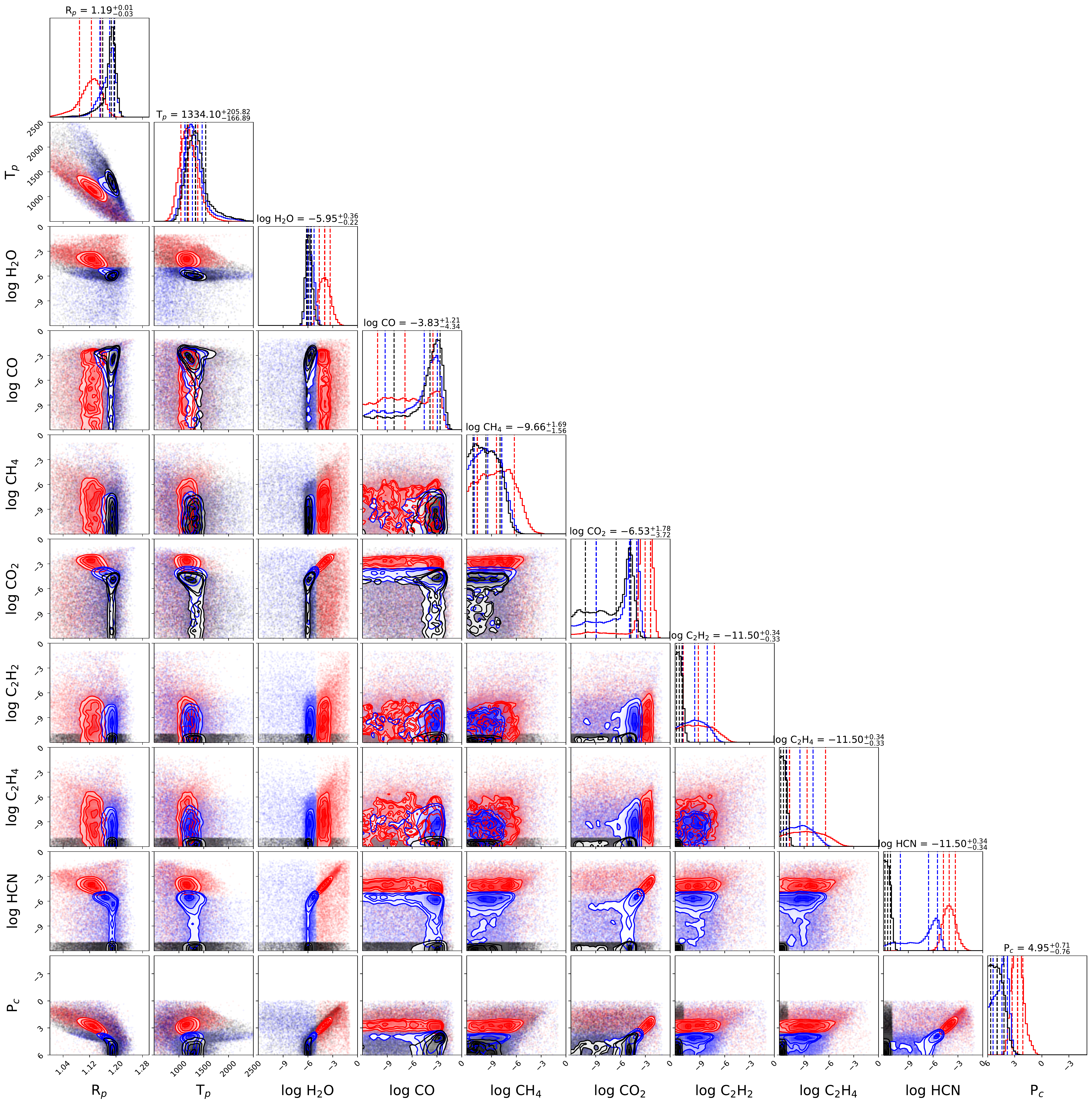}
    \caption{Posterior distributions from our different free retrieval scenarios. Black: Base retrieval; Red: Extended retrieval, solution 1; Blue: Extended retrieval, solution 2.}
    \label{fig:post_compa}
\end{figure*}

More precisely, our `base' setup detected the presence of water, carbon monoxide and carbon dioxide. The particularly high signal-to-noise ratio in this dataset allows to extract precise constraints on the abundances of these molecules. Our Bayesian analysis (see black posterior distribution in Figure \ref{fig:post_compa}) found the abundance of water to be log(H$_2$O) = -5.95$^{+0.36}_{-0.22}$, the abundance of carbon monoxide to be log(CO) = -3.83$^{+1.21}_{-4.34}$ and the abundance of carbon dioxide to be log(CO$_2$) = -6.53$^{+1.78}_{-3.72}$. We note that the recovered water abundance is low, which does not match the expected predictions from theoretical models and other derived abundances from similar planets \citep{Tsiaras_pop_study_trans, sing_exoplanets, Pinhas_ten_HJ_clouds}.  \\

From the breakdown of the contributions (Bottom part of Figure \ref{fig:spectrum}), we deduce that KELT-11\,b presents a strong water signature, well defined by the 1.4$\mu$m feature, which leads to a very accurate estimation of its abundance. With respect to CO$_2$ and CO, the signal of these molecules is weaker in the HST wavelength range and possess some similarities. This degeneracy leads to a larger margin of error for the two molecules. However, the additional absorption from 1.4$\mu$m to 1.6$\mu$m clearly indicates that this model requires a combination of these two molecules. In addition to these three molecules, we also constrain the abundance of methane to be lower than log(CH$_4$) $\lesssim$ -8 to 1$\sigma$. The updated list of parameters for the atmosphere and the orbit of KELT-11\,b is summarised in Table \ref{tab:evidence_compa}.

The recovered temperature is about 1300K, which is expected since the observations are probing the terminator region, which is naturally colder than the day side of the planet, better represented by the equilibrium temperature of about 1700K. Many studies pointed out that these observed differences between equilibrium and terminator temperatures, which are observed for almost all planets \citep{Tsiaras_pop_study_trans, Skaf_2020}, could be caused by 3-dimensional biases not accounted for in 1-dimensional models \citep{Feng_2016,caldas_3deffects, Irwin_w43b_phase, Pluriel_2020_biases,MacDonald_2020_cold, changeat_2020_phase1, feng2020_2d,changeat_2020_phase2}.

Contrary to suggestions by \cite{Zak_kelt11_2019}, the retrieval presented here indicates a relatively clear atmosphere and does not recover evidence of high altitude clouds. We constrain the top pressure for the clouds to be P$_c$ $\gtrsim$ 0.1 bar. These results could indicate that the atmosphere of KELT-11\,b is depleted in sodium and lithium. Other possibilities include a hazy atmosphere, with more opaque absorption at lower wavelengths.

%The run presented in the results section and described above did not require any sort of fine tuning, ensuring that these results are robust and directly express the information content in the spectrum. showing abundant quantities of CO and CO$_2$ but low amount of CH$_4$ (see posterior upper limit constrain in Figure \ref{fig:post}),

Since methane shares similar features to H$_2$O, we investigate whether the 1.4$\mu$m signal could be from this molecule. Forcing methane to an abundance higher than 10$^{-7}$ highlighted that the stronger absorption shape of methane in the lower (1.2$\mu$m) wavelengths (compared to the main 1.4$\mu$m feature), as well as the tighter absorption at 1.4$\mu$m, do not match the spectrum. To explore the significance of this model, we ran several other scenarios that are presented in the next sections.

\subsection{Extended model results}

In the `extended' model, we add HCN, C$_2$H$_2$ and C$_2$H$_4$, which allows us to explore a wider range of carbon bearing species. As seen in the previous section, the shape of the spectrum at 1.5$\mu$m is well fit with CO and CO$_2$. However, HCN also shows strong features at 1.5$\mu$m, which could help the retrieval to fit the observed additional absorption at these wavelengths.

The `extended' retrieval unveiled two solutions that are highlighted respectively in red and blue: Figure \ref{fig:spectrum_compa} for the spectra and Figure \ref{fig:post_compa} for the posterior distributions.

As opposed to the `base' solution, Solution 1 does not contain a high abundance of CO but presents better constrained posteriors for CO$_2$ and HCN, with respective abundances of -3.1$^{+0.7}_{-0.9}$ and -4.1$^{+0.7}_{-0.7}$. The atmosphere is also consistent with the presence of opaque clouds and the recovered water abundance is higher than in the previous run: -4.0$^{+0.7}_{-0.7}$. The breakdown of the contribution from the different molecules for this run can be found in Appendix 3. It shows where the additional HCN opacity contributes and illustrates the degeneracies between CO, CO$_2$ and HCN. Solution 2 is similar to what was found in the previous section with the `base' model, when it comes to H$_2$O, CO and CO$_2$. The atmosphere is consistent with absorption of H$_2$O and a mix of carbon bearing species without clouds. There are strong degeneracies between CO, CO$_2$ and HCN, where combinations of the 3 molecules can lead to statistically equally valid solutions. In the posterior distribution of those three molecules, the tails towards the lowest abundances are strong, meaning that their individual detection cannot be fully confirmed. In all runs, the C$_2$H$_2$ and C$_2$H$_4$ do not contribute to the fit and a upper limit of about 10$^{-6}$ is inferred. The recovered temperature is well constrained and remains similar, in both `base' and `extended' solutions, at around 1300 K.

\subsection{Water only retrieval results}

\begin{figure}
    \centering
    \includegraphics[width=\columnwidth]{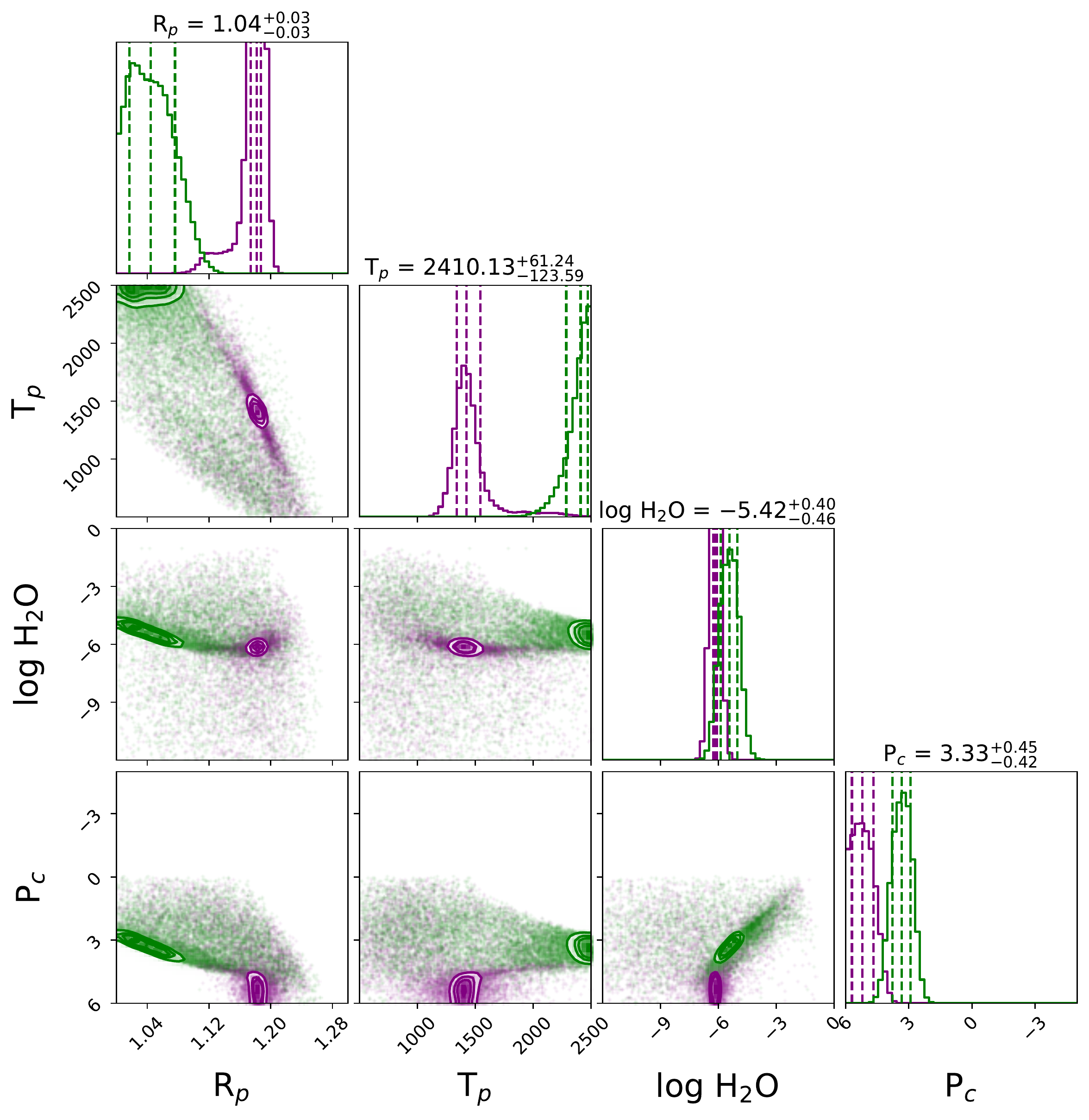}
    \caption{Posterior distributions from the `water only' scenarios. Purple: water only; Green: water only (no CIA).}
    \label{fig:post_water_only}
\end{figure}

In order to test the need for CO, CO$_2$ and HCN, we perform a complementary retrieval without those molecules (water only model). The best-fit spectrum is compared with the other runs in Figure \ref{fig:spectrum_compa} (purple), while the posterior distribution is presented in Figure \ref{fig:post_water_only}. 

Results from this run are particularly interesting as they highlight the fact that the `water only' model is also a good fit of the observed spectrum. We note that the `water only' model only provides a good fit when CIA is included (when CIA is not included we get $\Delta$ log E = 58.7). The model `water only (no CIA)' is shown in green in Figure \ref{fig:spectrum_compa} and \ref{fig:post_water_only}. In the `water only' run, the water abundance drops to log(H$_2$O) = -6.2$^{+0.1}_{-0.1}$, which is much lower than solar abundances and might be nonphysical. The low abundance of water allow CIA to become dominant at longer wavelengths and provide a good fit to the observed signal after 1.5$\mu$m. The breakdown of the contribution from the different species for the `water only' run is presented in Appendix 3.

\subsection{Full retrieval}

In our `full' retrieval we also consider the absorption from the optical absorbers: TiO, VO and FeH. Evidence for these has been presented in a number of previous HST G141 transmission studies \citep[e.g.][]{edwards_w76,Skaf_2020}. The obtained log E is higher than with previous models, however it still remains in the same range of evidence: the difference with the `water only' model is only 2.4. The spectrum is highlighted in maroon in Figure \ref{fig:spectrum_compa}, while the full posterior distribution for this model can be found in Appendix 4. As shown in the posterior distribution, the additional flexibility provided by the optical absorbers leads to a different solution. Here, it shows that the HST spectrum can be fit with a much higher abundance of water (log H$_2$O = 3.6$^{+0.6}_{-0.7}$) associated with log TiO = -5.1$^{+0.6}_{-0.8}$. This water mixing ratio is consistent with previously observed abundances for this type of planets \citep{Tsiaras_pop_study_trans} but the presence of TiO in this atmosphere would be surprising given the recovered temperature of about 1300K. Additionally, this solution requires a high abundance of CO$_2$ and HCN (both around 10$^{-3}$), while CO cannot be confirmed. It is important to notice that, as demonstrated with the `extended' runs, the abundances for CO, CO$_2$ and HCN are mainly driven by the points after 1.5$\mu$m. They thus present degeneracies and could be sensitive to random scatter of those data points. To properly confirm those abundances, other observations covering an additional independent feature of each of those molecules would be required, which could be provided by the next generation of telescopes. In this run, an opaque cloud cover is preferred in this solution, which might be compatible with the findings from \cite{Zak_kelt11_2019}.

\section{Discussion}

As seen in the previous section, the example of KELT-11\,b highlights how the choice of model can lead to different solutions and interpretations when performing atmospheric retrieval studies. In addition to this, when statistical methods are employed for model selection in HST data (e.g: comparison of Bayesian evidence), a better statistical evidence does not always guarantee that the favoured model is the right one, since other unexplored scenarios might also provide a decent fit. This issue is a strong indication that retrieval studies should attempt to assess the information content in exoplanet spectra by exploring a wide enough range of scenarios. In our exploration, water vapor is robustly detected by all models, with sub-solar abundances. As the `water only' model provides a good fit to this spectrum, following the law of parsimony or `Occam's razor' principle, one would be tempted to prefer this model. However, from a physical and chemical perspective, this model would imply that the atmosphere has an extremely low amount of water and is depleted in all other absorbing molecules in the wavelength considered. In this case, a model of higher complexity might provide a better description for this atmosphere. When carbon species are added, CO$_2$ is recovered with abundances that variate between 10$^{-7}$ and 10$^{-3}$. CO and HCN are detected, depending on the choice of model and with large posterior tails in the `base' and `extended' models. Those tails might be explained by the fact that the water only model is already a good fit and by the similitude in the CO, CO$_2$ and HCN contributions for the considered wavelength range. This could also indicate that the detection of carbon bearing species might be subject to systematic errors in the HST data, similar to what was found for VO in the case of WASP-121\,b \citep{Evans_Wasp121b_spectrum_em, Mikal_Evans_2019}. A larger wavelength coverage from additional observations might help resolve this issue.

In addition to model dependant solutions, other aspects of retrieval studies could lead to different interpretations. In the first place, one can investigate the impact of using self-consistent schemes to represent chemical abundances. Furthermore, in the search for more precise characterisations, it is common to combine instrument observations, which in theory increase the information content on which to retrieve. However, such method can introduce systematic errors that should be investigated in the case of the TESS+HST data of KELT-11\,b.

\subsection{Comparison with an equilibrium scheme}

\begin{figure}
    \centering
    
    \includegraphics[width=\columnwidth]{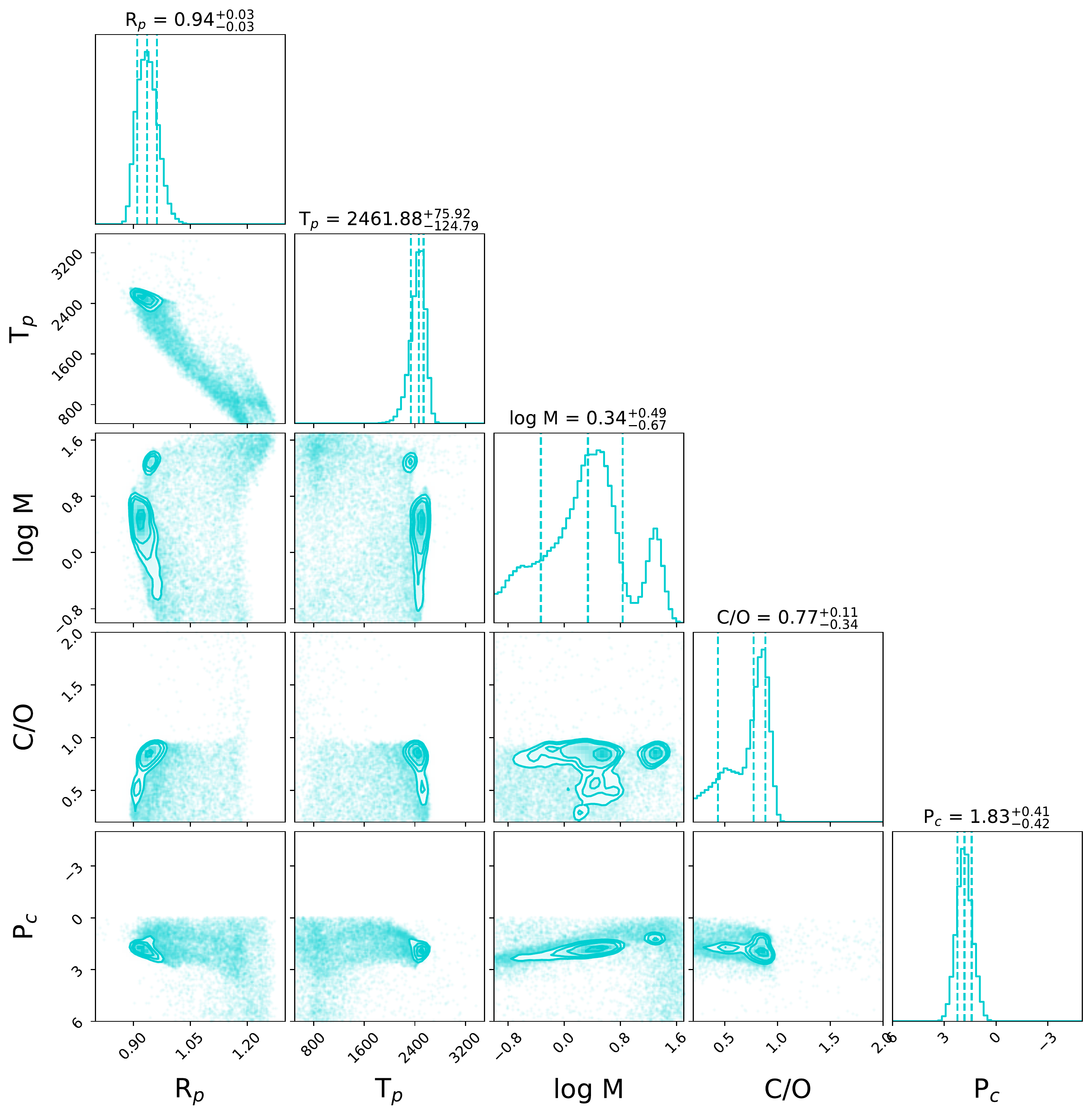}
    \includegraphics[width=\columnwidth]{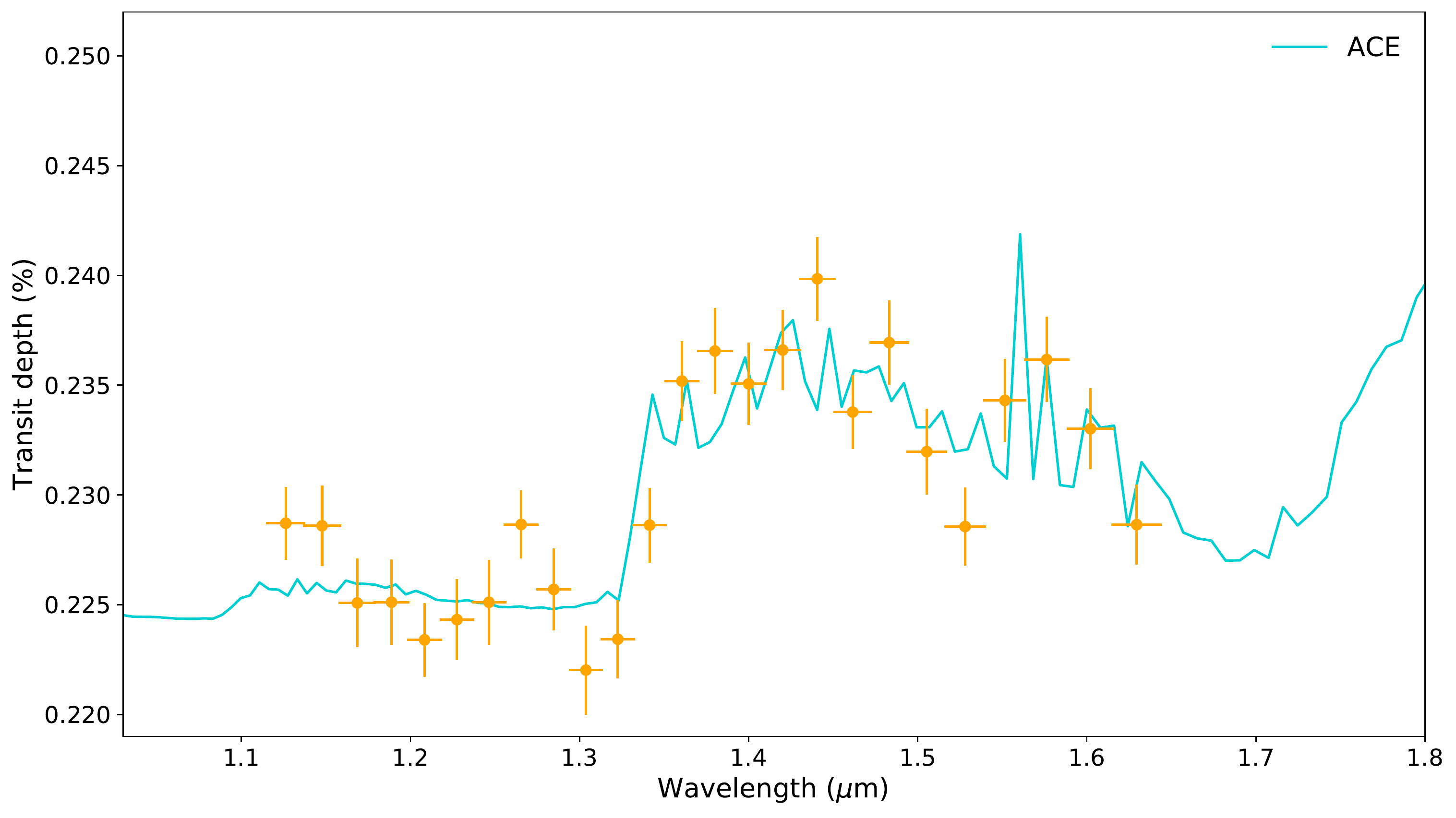}
    \caption{Posterior distribution (top) and best fit spectrum (bottom) from the `ACE' scenario. This run assumes equilibrium chemistry using the scheme from \cite{Agundez_2dchemical_HD209_HD189}.}
    \label{fig:post_eq}
\end{figure}

Since HST observations have low information content, which typically only allow H$_2$O to be constrained, equilibrium chemistry models are also convenient to extrapolate the behaviour of the other molecules. Importantly, their implied assumptions on the state of the planet and its physical/chemical behaviour often neglect phenomena of major importance such as 3-dimensional effect, dynamical effects, disequilibrium processes, to only name known sources of biases \citep{2015_venot_diseq,caldas_3deffects,Drummond_2020_coupled,Changeat_2019_2l, Changeat_2020_alfnoor}. As the physics of such systems can be extremely complex and far from any environment we know in the Solar System, the selection of a particular chemical model may lead to results biased by preconception. Nevertheless, it seems from our free exploration of the planet that carbon species might play an important role in shaping the WFC3 transmission spectrum of KELT-11\,b, which could provide robust constraints to investigate equilibrium chemistry schemes for this planet. The posterior distribution and spectrum from our equilibrium chemistry run can be found in Figure \ref{fig:post_eq}.

\begin{figure}
    \centering
    \includegraphics[width=\columnwidth]{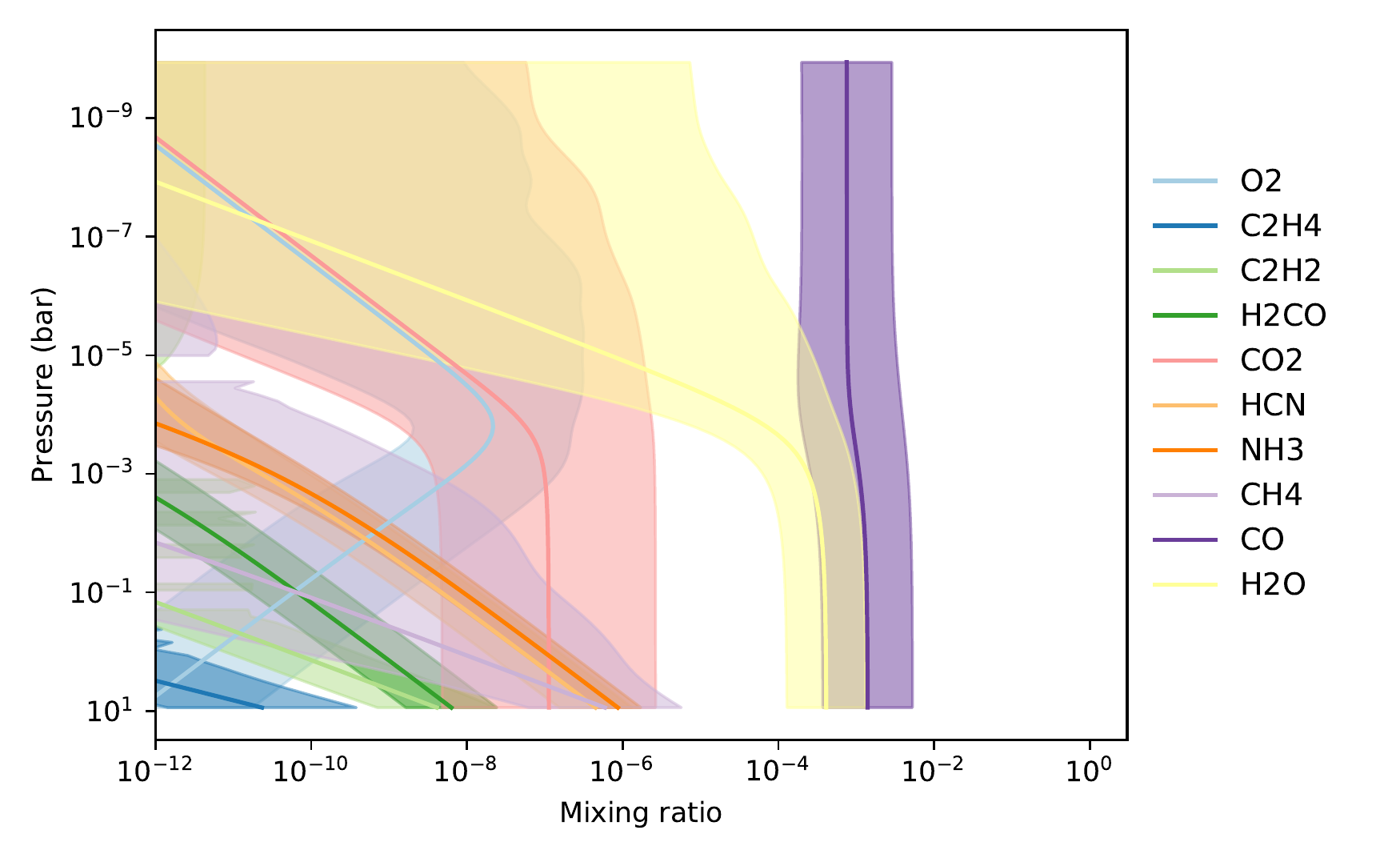}
    \caption{Mixing ratio with altitude of the active species in our `ACE' scenario.}
    \label{fig:eq_abundances}
\end{figure}

For a more direct comparison, we also display the recovered abundances with altitude for this run in Figure \ref{fig:eq_abundances}. As can be seen in this figure, the dominant species in this atmosphere are H$_2$O, CO and CO$_2$, thus confirming the pertinence of our `base' scenario. We however note that the additional constraints from the assumption of equilibrium chemistry lead to higher abundances for those molecules, as can be inferred from the high retrieved metallicity of log M = 0.3$^{+0.5}_{-0.7}$. The water abundance here is about 10$^{-3}$, which is closer to solar abundances than the free results. Other instances of high metallicity atmospheres have already been observed in exoplanets \citep[e.g.][]{Wakeford_2017,spake2019supersolar, MacDonald_2019}. A noticeable point is that the contribution from the more exotic carbon species added in the `extended' runs (HCN, C$_2$H$_2$ and C$_2$H$_4$) remain minor when the equilibrium chemistry retrieval is used. Their recovered abundances are below 10$^{-6}$. Finally, the recovered temperature becomes about 2400K, which is higher than expected for this planet and might reflect remaining biases in this retrieval, especially as we can observe some evident correlations with other parameters. For comparison, we obtain $\Delta log E = 59.3$, which is much lower than any of the investigated free runs but help providing a sense of what one should expect in such planet. This lower log E might however provide evidence that the assumption of equilibrium chemistry does not hold for this atmosphere.

\subsection{Impact of TESS in retrieval analyses}

Motivated by the narrow wavelength coverage of the HST-G141 grism, many studies attempt to combine with other instruments, either adding HST-STIS, Spitzer, TESS and/or ground based observations \citep[e.g.][]{sing_exoplanets}. In particular, Spitzer covers the CH$_4$ and the CO/CO$_2$ absorption bands with the photometric channels at 3.6$\mu$m and 4.5$\mu$m. TESS and HST-STIS cover shorter wavelengths, which are particularly sensitive to clouds/hazes properties and absorption from atomic molecules and metal hydride/oxides (TiO, TiH, VO, FeH to name a few). However recent studies highlighted the danger of performing such combinations without investigating potential incompatibilities with the datasets \citep{Yip_lightcurve,pluriel_2020_ares,yip2020compatibility}. In particular, issues can arise when datasets are reduced using different orbital parameters \citep{Yip_lightcurve, alexoudi_inc} or limb darkening coefficients \citep{Tsiaras_pop_study_trans}. Furthermore, stellar variability and activity can produce spectra offsets when observations are not taken simultaneously \citep{Bruno_2019}. Finally, imperfect correction of instrument systematics can lead to inconsistent results \citep{stevenson_w43_1,stevenson_w43_2,Diamond_Lowe_2014}. 

\begin{figure}
    \centering
    \includegraphics[width=0.95\columnwidth]{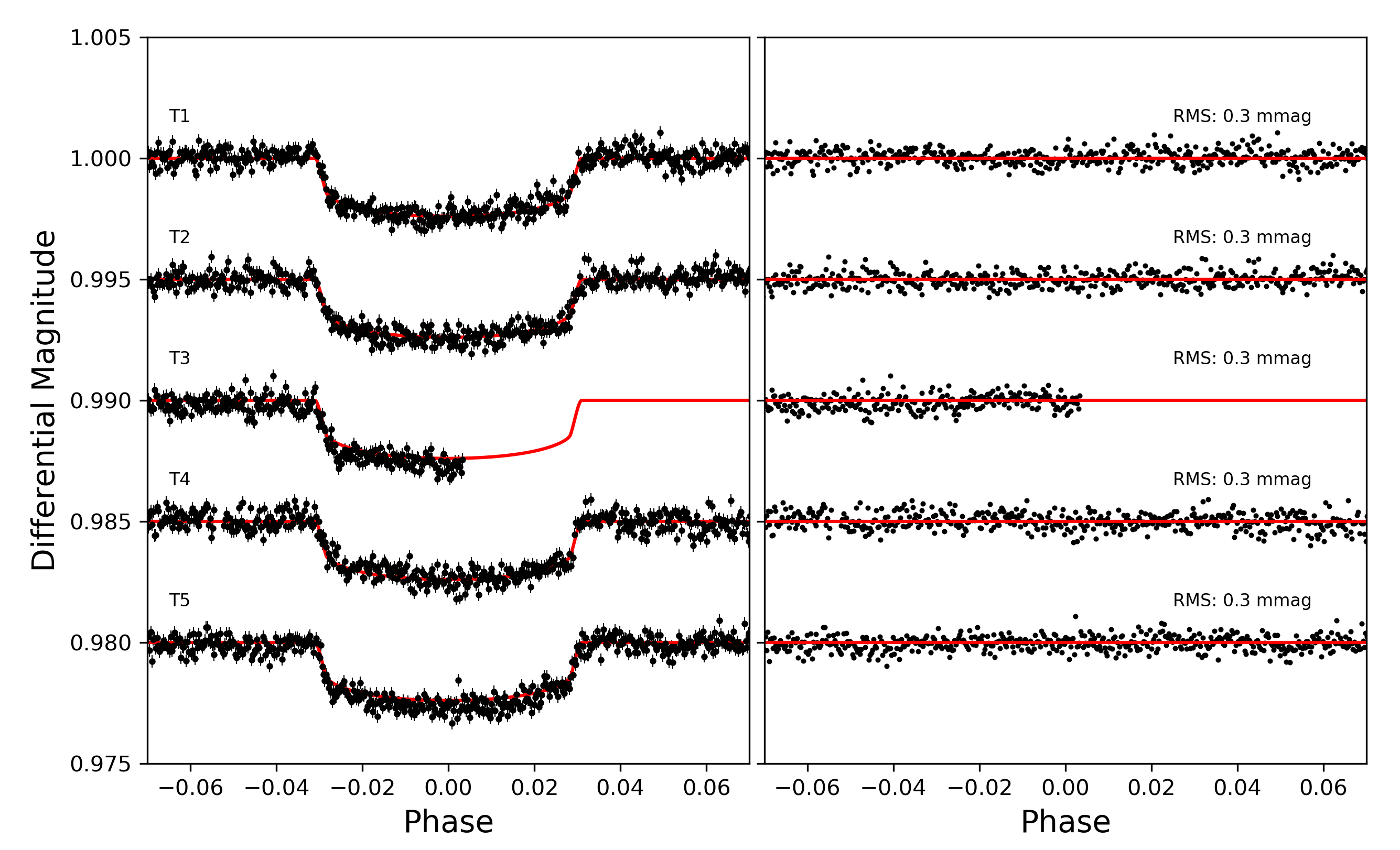}
    \includegraphics[width=\columnwidth]{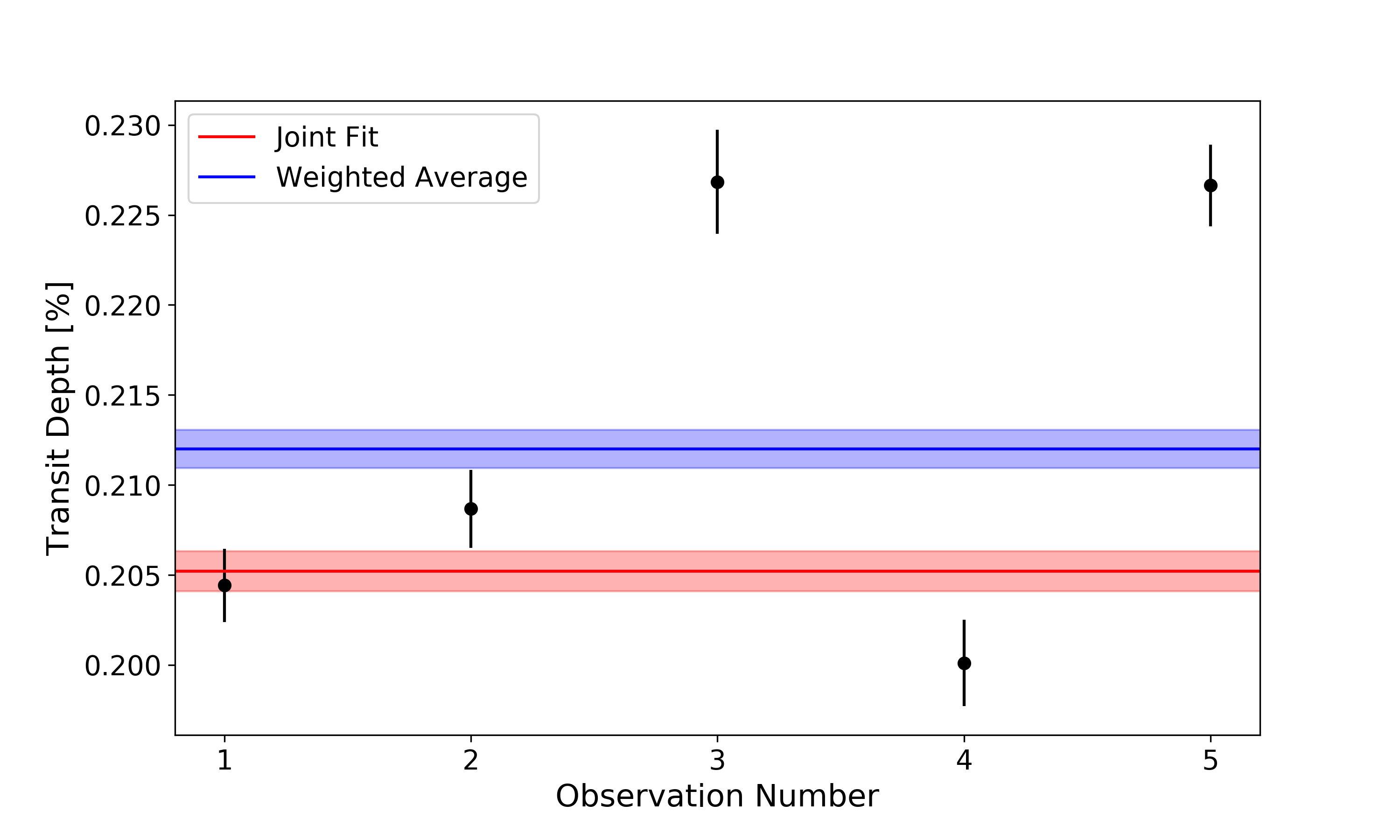}
    \caption{Top: Fitting of the TESS observations presented in this work. Left: detrended data and best-fit model. Right: residuals from fitting. Bottom: Recovered TESS depths for each transit (black) and from a joint fit of all the observations (red). The transit depth is seen to vary drastically. The weighted average of the individual fits is also shown (blue) and disagrees with the joint fit.}
    \label{tess_depth}
\end{figure}

As explained previously, for KELT-11\,b we discard entirely the available Spitzer data, due to the missing pre-ingress part of the transit. We however considered the addition of the TESS data since 5 transits were observed, making sure the orbital parameters and limb-darkening coefficients are consistent with the HST-G141 grism reduction. In Figure \ref{tess_depth}, we plot the recovered TESS depths for each transit. As can be seen in this figure, there are large variations of the observed depths between the different transit.

These differences could come from variability in the environment (e.g. stellar activity, observation contamination) and/or imperfectly corrected systematics from the reduction. This leads to large differences whether we choose to combine the observations using a joint-fit or a weighted average, confirming the difficulty of using TESS for atmospheric characterisation in this case. We note that the third observation only covered half of the transit which may explain the discrepancy of this particular fit. When performing a retrieval analysis of a combined TESS+HST dataset, many optical absorbers would have their posterior strongly defined by the TESS data, which could lead to biased results. In addition to this, when different reduction pipelines are used for HST-G141, the same spectrum shape is usually obtained but it is common to observe differences in the white light curve depth. Such an offset would only translate into biases for the radius when considering the HST data alone, but could lead to unstable results when combining with other instruments. 

Fully aware of these potential incompatibilities, we added the obtained TESS joint-fit/weighted averaged and performed an atmospheric retrieval similar to our `full' scenario. The results from those retrievals, which are available in Appendix 5, highlight possible incompatibilities between the two datasets, as discussed above. Both solutions, from the joint-fit and the weighted averaged TESS data, lead to nonphysical solutions with high abundances of H$_2$O and either HCN or CO$_2$. It is interesting to see that the low observed TESS photometric point leads to the replacement of TiO by FeH in those runs. Those retrievals are good examples illustrating why extra care might be required when combining instrument observations in retrieval studies. In the case of KELT-11\,b, the datasets may have been made incompatible due to uncorrected systematics in the HST data as the white light curve residuals seen here, and in \cite{coln2020unusual}, are non-gaussian.

To understand if there is indeed an offset between these datasets, further data is required. While STIS data has not been acquired for this planet, G102 will soon be obtained (PN: 15926, PI: Knicole Colon, \citet{more_data_to_steal}) and provide further insights into the nature of this planet, helping to distinguish between the potential compositions presented here.

\subsection{Comparison with other literature results}

The same dataset from the HST G141 grism was recently analysed in an independent study from \cite{coln2020unusual}. While different reduction pipelines were used, their main conclusions are similar to ours, unveiling a spectrum with similar spectral shape to what is presented here. The studies strongly agree on the presence of water vapour in this planet, with sub-solar abundances that vary depending on model assumptions. Due to the shape of the spectrum, the need for an additional absorption after 1.5$\mu$m is also highlighted. In \cite{coln2020unusual}, this is attributed to HCN. In our study, we show this could be well matched by a mix of carbon bearing species, which include HCN, but we were not able to statistically attribute the features to this molecule only (see Table \ref{tab:evidence_compa}). The evidence for TiO is highlighted in \cite{coln2020unusual}, which is strongly confirmed by their addition of the TESS data. However, as mentioned in the previous section, the combination of datasets from different instruments with no wavelength overlap should be done carefully. In particular, our spectrum, while having a similar spectral shape, is offset by about 200 ppm to the one used in \cite{coln2020unusual}. Similar offsets of about 400 ppm are also observed in their paper as well, when analysing the spectra obtained with different pipelines. In our `full' retrieval, which includes TiO as well as other near-visible species, we find that TiO is also favoured in the HST dataset (this is driven by the data points around 1.3$\mu$m), but when the TESS data is added, TiO disappears in favour of FeH, due to the higher HST spectrum recovered in our study. Thus, identification of a near-visible absorber and/or its precise abundance is likely to be biased by systematic offsets when TESS+HST is used. Finally, most of our retrievals recover a well defined temperature around 1300K, which is expected from previous transit observations \citep{Tsiaras_pop_study_trans, caldas_3deffects, Pluriel_2020_biases, MacDonald_2020_cold, Skaf_2020}. However, this disagrees with the findings from \cite{coln2020unusual}, which find a lower limit on the temperature of about 1900 K. These differences could either be due to the use of different reduction/retrieval pipelines or differences in the choice of models (temperature/clouds).

\section{Conclusion}

Having one of the highest signal-to-noise ratios of the known exoplanets and a super extended atmosphere, KELT-11\,b will be a prime target for future observatories. We analysed the HST G141 spectrum of the planet KELT-11\,b. From our spectral retrieval exploration, we confirmed the presence of  water vapour, with sub-solar abundances. In addition to this, the rich spectrum features an additional absorption after 1.5$\mu$m, which is clearly detected in all our retrieval scenarios. This could come from a mixture of carbon bearing species (CO, CO$_2$ or HCN) and while equilibrium chemistry seem to favor CO and CO$_2$, the spectrum does not contain enough information to clearly identify the mix of compounds. However, when included, CO$_2$ seems to be systematically detected, with varying abundances depending on the model. The high abundance of carbon species, inferred from the base model, and the relatively low abundance of water suggest a planet with high C/O ratio \citep{2015_venot_diseq}. This could have important implications regarding the formation processes for this planet and potential inform formation and evolution models \citep{_oberg_2011_planetforma}. A high C/O ratio, along with the contrast between our rich spectrum and the data from the ground at shorter wavelengths \citep{Zak_kelt11_2019}, showcases a planet with particularly interesting physics and chemistry. Observations with future observatories, such as the James Webb Space Telescope \citep{greene_jwst}, Twinkle \citep{Edwards_twinkle} and Ariel \citep{Tinetti_ariel}, would dramatically enhance our comprehension of this world and thus provide outstanding information for chemical models and formation theories.

Using this planet as an example, we explored model dependant behaviour in retrieval analysis and highlighted the dangers of assuming a particular physics (self-consistent schemes) when trying to extract information content from exoplanet spectra. We also experimented the behaviour of retrieval analyses on combined dataset, by adding the available TESS data to our HST spectrum of KELT-11\,b.

\section*{Acknowledgements}

This project has received funding from the European Research Council (ERC) under the European Union's Horizon 2020 research and innovation programme (grant agreement No 758892, ExoAI) and under the European Union's Seventh Framework Programme (FP7/2007-2013)/ ERC grant agreement numbers 617119 (ExoLights). Furthermore, we acknowledge funding by the Science and Technology Funding Council (STFC) grants: ST/K502406/1, ST/P000282/1, ST/P002153/1 and ST/S002634/1.

We acknowledge the availability and support from the High Performance Computing platforms (HPC) DIRAC and OzSTAR, which provided the computing resources necessary to perform this work.

This work is based on observations made with the NASA/ESA Hubble Space Telescope. These publicly available observations were taken as part of proposal 15225, led by Knicole Colon \citep{colon_proposal}. These were obtained from the Hubble Archive which is part of the Barbara A. Mikulski Archive for Space Telescopes. Additionally, this paper includes data collected by the TESS mission which is funded by the NASA Explorer Program. TESS data is also publicly available via the Barbara A. Mikulski Archive for Space Telescopes (MAST). We are thankful to those who operate this archive, the public nature of which increases scientific productivity and accessibility \citep{peek2019}.\\

\textbf{Software:} corner \citep{corner}, Iraclis \citep{Tsiaras_2016_iraclis}, TauREx3 \citep{al-refaie_taurex3}, pylightcurve \citep{tsiaras_plc}, ExoTETHyS \citep{morello_exotethys}, Astropy \citep{astropy}, h5py \citep{hdf5_collette}, emcee \citep{emcee}, Matplotlib \citep{Hunter_matplotlib}, Multinest \citep{Feroz_multinest}, Pandas \citep{mckinney_pandas}, Numpy \citep{oliphant_numpy}, SciPy \citep{scipy}.

\bibliographystyle{aasjournal}
\bibliography{main}

%%%%%%%%%%%%%%%%%%%%%%%%%%%%%%%%%%%%%%%%%%%%%%%%%%%%%%%%%%%%%%%%%%%%%%%%%%%%%%%%

%\newpage
\renewcommand{\floatpagefraction}{.9}%
% \section{APPENDIX}
%\appendix 

\appendix

\section*{Appendix 1: Parameters and limb-darkening coefficients used in this work}

\begin{table*}[h]
    \centering
    \begin{tabular}{ccc}
    \hline\hline
    {\bf Parameter} & {\bf Value} & {\bf Unit} \\\hline\hline
    \multicolumn{3}{c}{Stellar parameters} \\
    Radius (R$_{\rm s}$) & 2.69 $\pm$ 0.22 & R$_{\odot}$ \\
    Mass (M$_{\rm s}$) & 1.44 $\pm$ 0.43 & M$_{\odot}$ \\
    Temperature (T$_{\rm s}$) & 5375 $\pm$ 25 & K \\
    Surface Gravity (log g) & 3.7 $\pm$ 0.1 & cgs \\
    Metallicity ([Fe/H]) & 0.17 $\pm$ 0.07 & - \\ 
    \hline
    \multicolumn{3}{c}{Orbital parameters} \\
    Transit Mid Time (T$_0$) & 2457483.431 $\pm$ 0.0007 & BJD$_{\rm TDB}$\\
    Period (P) & 4.73613 $\pm$ 0.00003 & days \\
    Inclination (i) & 85.3 $\pm$ 0.3 & degrees \\
    Semi-major axis to star radius ratio (a/R$_{\rm s}$) & 4.98 $\pm$ 0.05 & - \\
    Eccentricity (e) & 0.0007$^{+0.002}_{-0.0005}$ & - \\
    \hline
    \multicolumn{3}{c}{Planet parameters} \\
    Radius (R$_{\rm p}$) & 1.35 $\pm$ 0.10 & R$_{\rm J}$ \\
    Mass (M$_{\rm p}$) & 0.171 $\pm$ 0.015 & M$_{\rm J}$ \\
 \hline\hline
    \end{tabular}
\caption{Parameters used in this work from \cite{Beatty_2017_kelt11}.}
\end{table*}

\begin{table}
    \centering
    \begin{tabular}{ccccc}\hline\hline
    Wavelength [$\mu$m] & c1 & c2 & c3 & c4\\\hline\hline
1.1153 - 1.1372	&	0.4729	&	-0.0342	&	0.2561	&	-0.1297	\\
1.1372 - 1.1583	&	0.4532	&	0.0230	&	0.2024	&	-0.1152	\\
1.1583 - 1.1789	&	0.4583	&	0.0242	&	0.1844	&	-0.1081	\\
1.1789 - 1.1987	&	0.4508	&	0.0545	&	0.1499	&	-0.0976	\\
1.1987 - 1.2180	&	0.4453	&	0.0679	&	0.1339	&	-0.0928	\\
1.2180 - 1.2370	&	0.4346	&	0.1111	&	0.0897	&	-0.0801	\\
1.2370 - 1.2559	&	0.4340	&	0.1239	&	0.0710	&	-0.0743	\\
1.2559 - 1.2751	&	0.4231	&	0.1646	&	0.0266	&	-0.0609	\\
1.2751 - 1.2944	&	0.4216	&	0.2388	&	-0.0943	&	-0.0180	\\
1.2944 - 1.3132	&	0.4060	&	0.2362	&	-0.0572	&	-0.0338	\\
1.3132 - 1.3320	&	0.4108	&	0.2483	&	-0.0812	&	-0.0252	\\
1.3320 - 1.3509	&	0.4035	&	0.2913	&	-0.1378	&	-0.0050	\\
1.3509 - 1.3701	&	0.4035	&	0.3216	&	-0.1822	&	0.0108	\\
1.3701 - 1.3900	&	0.4151	&	0.3198	&	-0.2017	&	0.0202	\\
1.3900 - 1.4100	&	0.4319	&	0.3159	&	-0.2135	&	0.0241	\\
1.4100 - 1.4303	&	0.4105	&	0.3725	&	-0.2713	&	0.0434	\\
1.4303 - 1.4509	&	0.4341	&	0.3518	&	-0.2740	&	0.0466	\\
1.4509 - 1.4721	&	0.4648	&	0.3302	&	-0.2903	&	0.0577	\\
1.4721 - 1.4941	&	0.5310	&	0.2144	&	-0.2116	&	0.0363	\\
1.4941 - 1.5165	&	0.5683	&	0.1675	&	-0.2020	&	0.0395	\\
1.5165 - 1.5395	&	0.6338	&	0.0681	&	-0.1479	&	0.0281	\\
1.5395 - 1.5636	&	0.6217	&	0.1085	&	-0.2146	&	0.0576	\\
1.5636 - 1.5889	&	0.6454	&	0.0502	&	-0.1806	&	0.0526	\\
1.5889 - 1.6153	&	0.6937	&	-0.0577	&	-0.0897	&	0.0221	\\
1.6153 - 1.6436	&	0.7903	&	-0.2637	&	0.0857	&	-0.0332	\\\hline\hline
    \end{tabular}
    \caption{Limb darkening coefficients used during light curve fitting.}
    \label{tab:limb_coeffs}
\end{table}{}

\clearpage

\section*{Appendix 2: Extracted spectrum for KELT-11\,b}

\begin{table*}[h]
    \centering
    \begin{tabular}{ccc}
    \hline\hline
    Wavelength [$\mu$m] & Transit depth [\%] & Error [\%] \\\hline\hline
1.1153 - 1.1372  & 0.22871 & 0.00161 \\
1.1372 - 1.1583 & 0.22859 & 0.00177 \\
1.1583 - 1.1789 & 0.22508 & 0.00196 \\
1.1789 - 1.1987 & 0.22511 & 0.00189 \\
1.1987 - 1.2180 & 0.22340 & 0.00162 \\
1.2180 - 1.2370 & 0.22432 & 0.00179 \\
1.2370 - 1.2559 & 0.22511 & 0.00188 \\
1.2559 - 1.2751 & 0.22866 & 0.00150 \\
1.2751 - 1.2944 & 0.22570 & 0.00181 \\
1.2944 - 1.3132 & 0.22202 & 0.00197 \\
1.3132 - 1.3320 & 0.22343 & 0.00173 \\
1.3320 - 1.3509 & 0.22862 & 0.00165 \\
1.3509 - 1.3701 & 0.23519 & 0.00177 \\
1.3701 - 1.3900 & 0.23655 & 0.00190 \\
1.3900 - 1.4100 & 0.23506 & 0.00182 \\
1.4100 - 1.4303 & 0.23660 & 0.00176 \\
1.4303 - 1.4509 & 0.23985 & 0.00185 \\
1.4509 - 1.4721 & 0.23378 & 0.00162 \\
1.4721 - 1.4941 & 0.23694 & 0.00186 \\
1.4941 - 1.5165 & 0.23197 & 0.00190 \\
1.5165 - 1.5395 & 0.22856 & 0.00172 \\
1.5395 - 1.5636 & 0.23431 & 0.00183 \\
1.5636 - 1.5889 & 0.23617 & 0.00189 \\
1.5889 - 1.6153 & 0.23302 & 0.00180 \\
1.6153 - 1.6436 & 0.22865 & 0.00176 \\ \hline\hline
    \end{tabular}
\caption{WFC3 transit depths and errors (in percent) for for KELT-11\,b.}
\label{tab:obs_spectrum}
\end{table*}
\clearpage

\section*{Appendix 3: Contributions of absorbing species to the best-fit models}

\begin{figure*}[h]
    \centering
    \includegraphics[width=0.8\columnwidth]{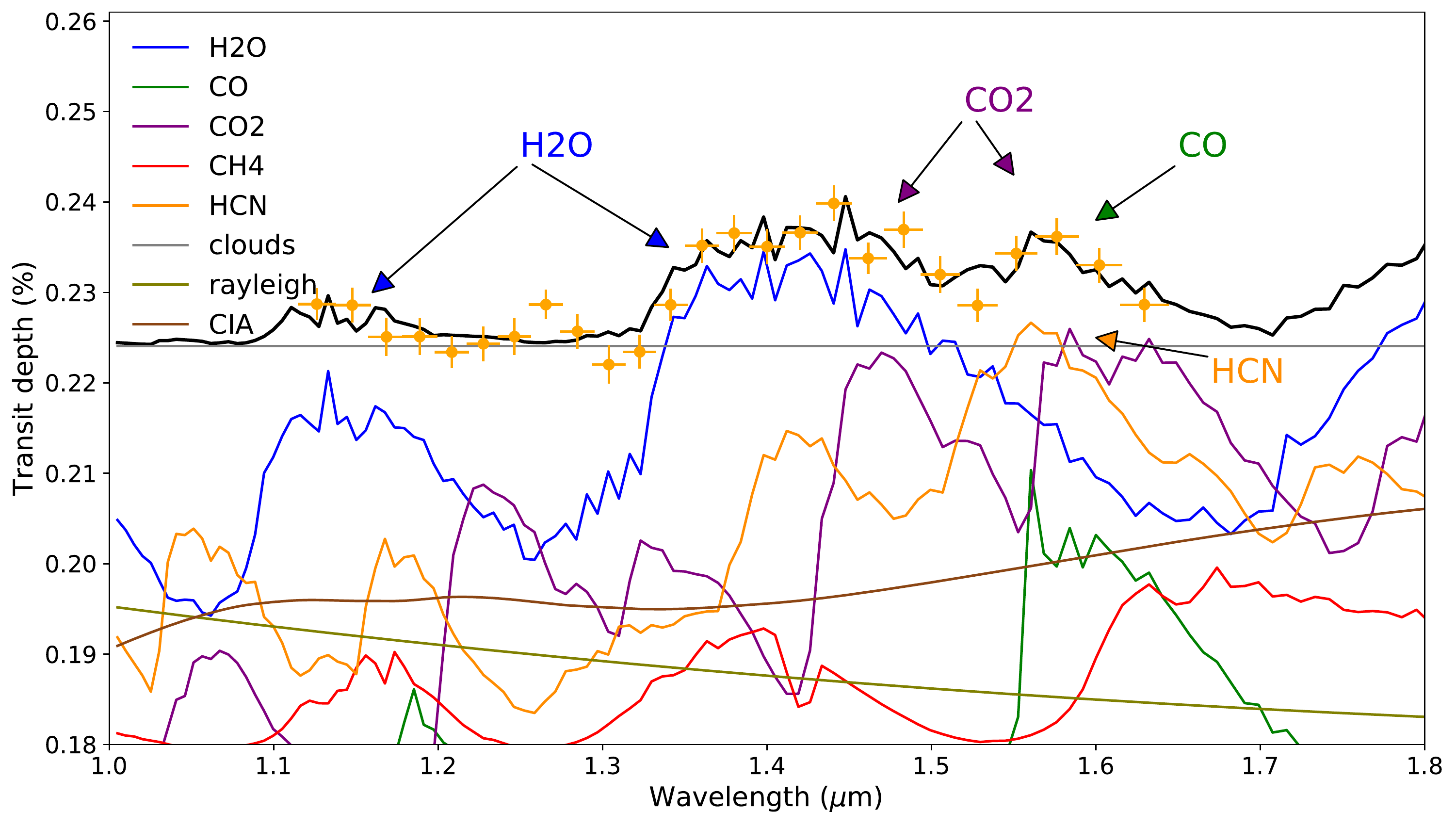}
    \includegraphics[width=0.8\columnwidth]{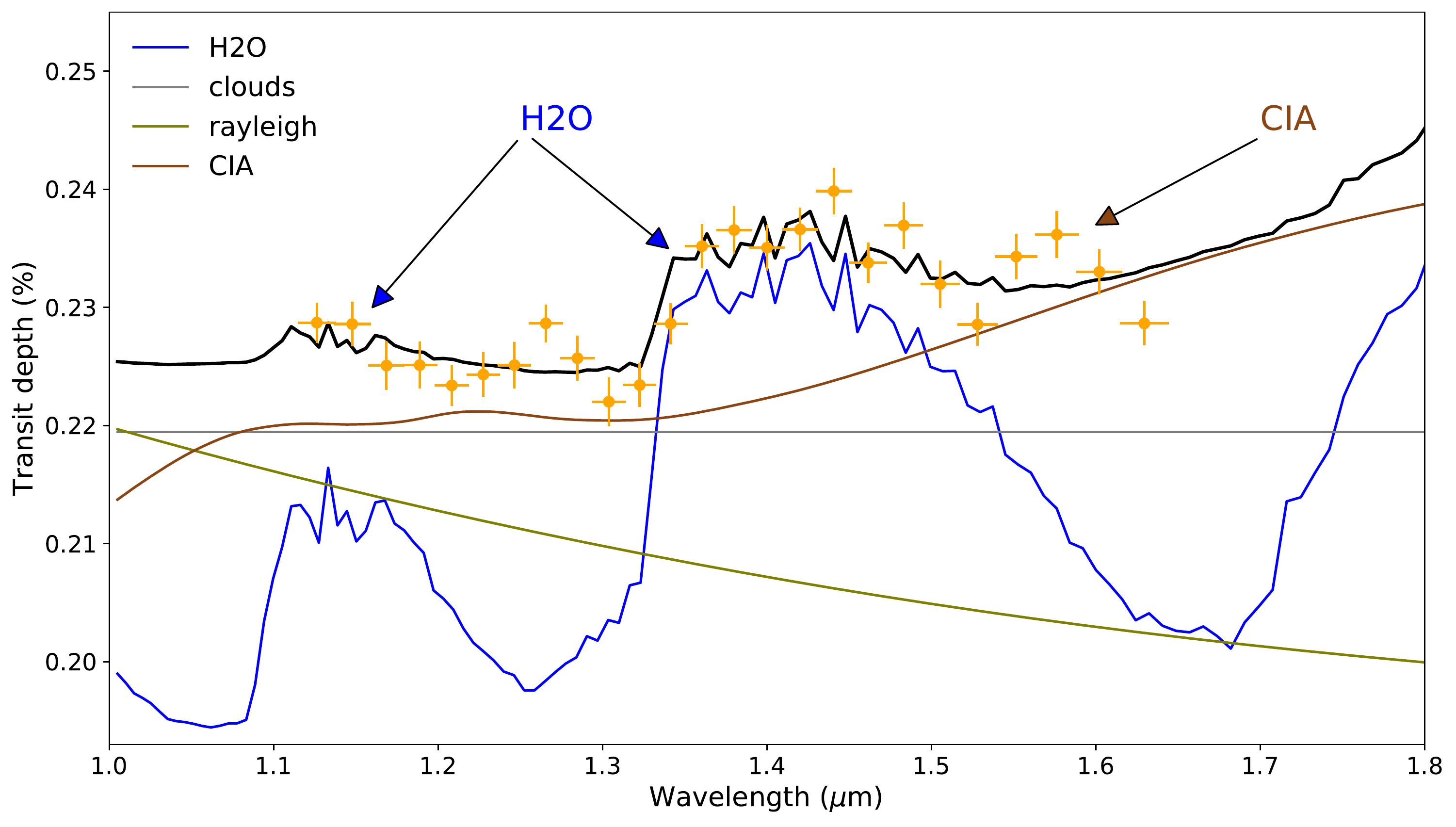}
    \caption{Best-fit contributions from the different absorbing species for the Solution 1 of the `extended' scenario (top) and the `water only' retrievals (bottom).}
    \label{fig:additional_contrib}
\end{figure*}

\clearpage

\section*{Appendix 4: Posterior distributions of the `full' model}

\begin{figure*}[h]
    \centering
    \includegraphics[width=0.9\columnwidth]{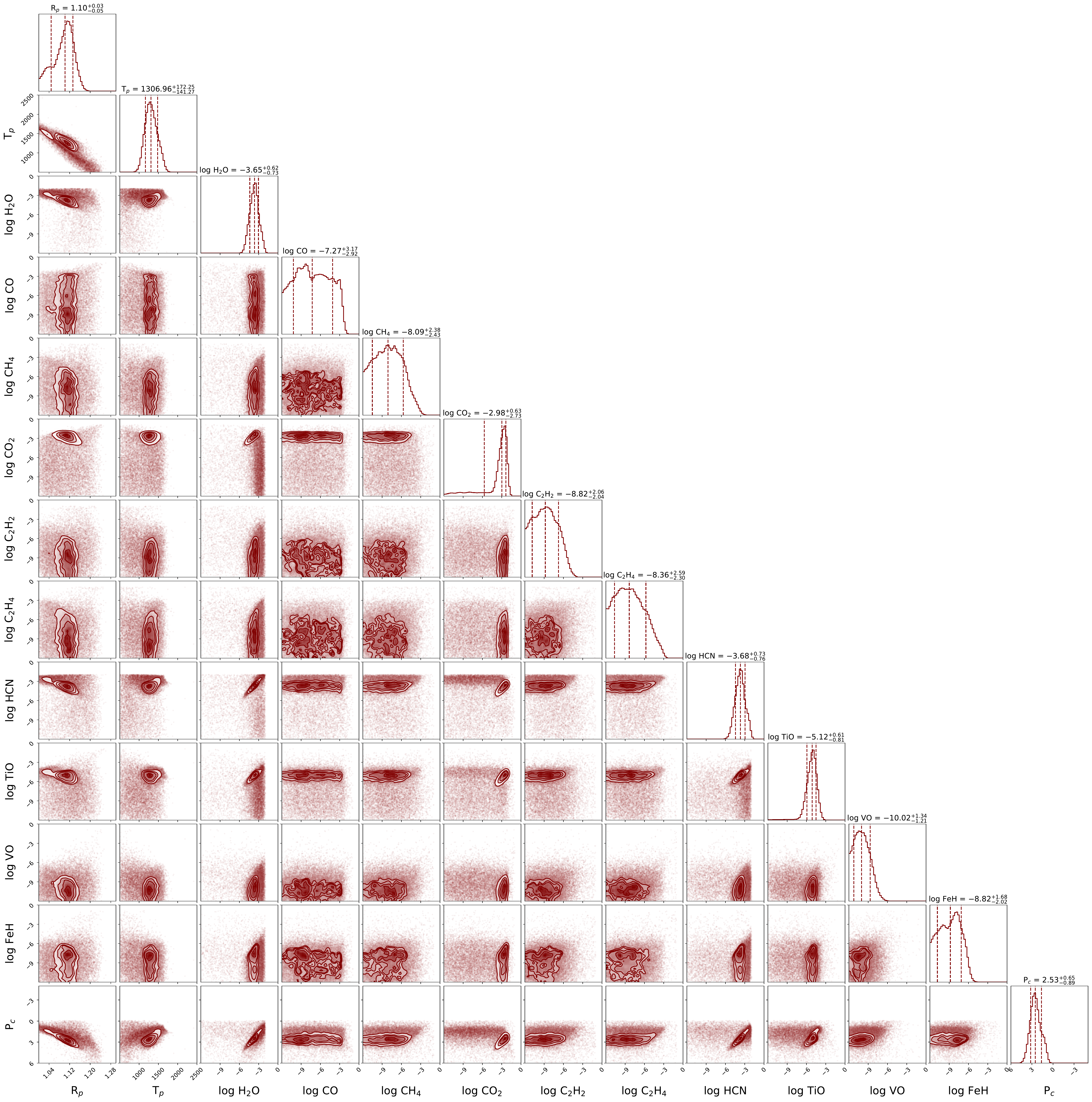}
    \caption{Posterior distribution of the `full' model, which includes the full range of carbon bearing species and the near-optical absorbers TiO, VO and FeH.}
    \label{fig:post_all}
\end{figure*}

\clearpage

%\begin{table*}
%    \centering
%    \begin{tabular}{ccc}
%    \hline\hline
%    Period (P) & 4.73620495$\pm$0.00000086 & days \\
%    Mid transit time (T$_{0}$) & 2458260.168608$\pm$0.000030 & BJD$_{TDB}$  \\
%    Inclination (i) & 85.3 $\pm$ 0.3 & degrees \\
%    Semi-major axis to star radius ratio (a/R$_s$) & 5.02 $\pm$ 0.07 & - \\
%    Eccentricity (e) & 0.002$^{+0.005}_{-0.0014}$ & - \\
 
% \hline\hline
%    \end{tabular}
%\caption{Summary of the updated parameters from our ephemeris refinement (HST + TESS).}
%\label{tab:recovered_params}
%\end{table*}

\section*{Appendix 5: Results of the combined retrievals}

\begin{figure*}[h]
    \centering
    \includegraphics[width=0.8\columnwidth]{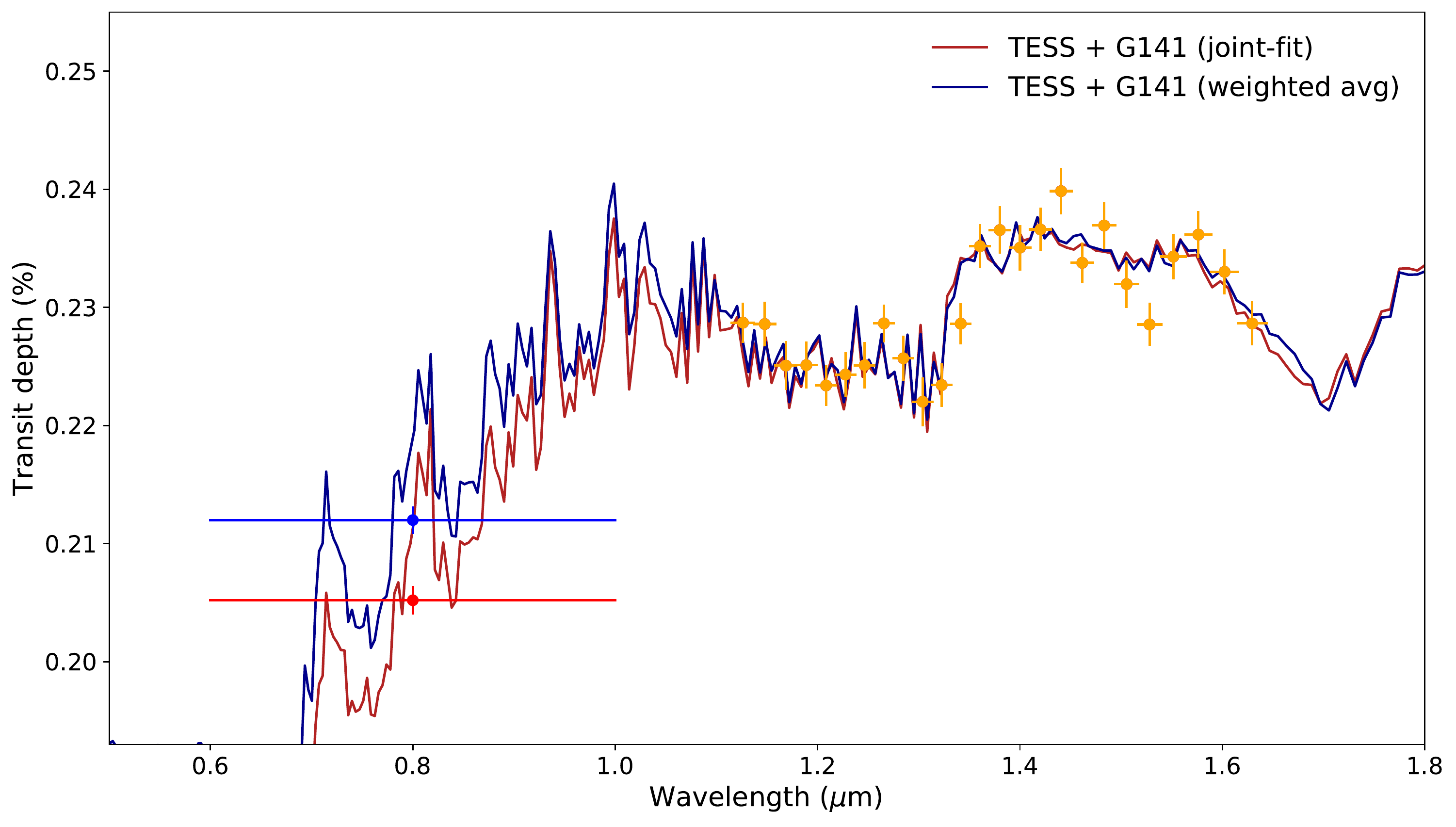}
    \caption{Best-fit spectra for the combined retrievals TESS+HST. Red: Retrieval using the joint-fit TESS data; Blue: Retrieval using the weighted averaged TESS data.}
    \label{fig:spectrum_tess}
\end{figure*}

\begin{figure*}[h]
    \centering
    \includegraphics[width=0.9\columnwidth]{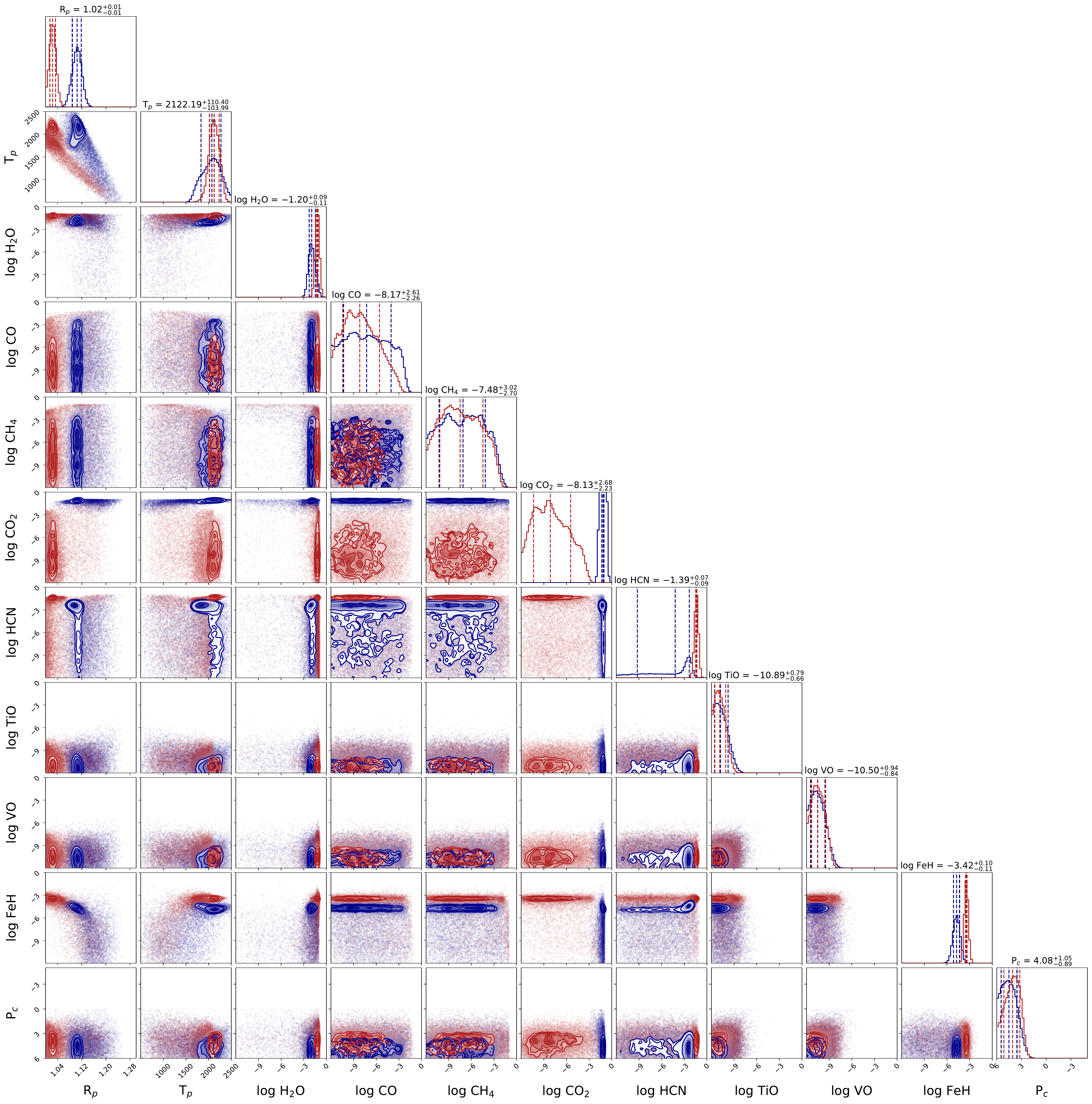}
    \caption{Posterior distribution for the combined retrievals TESS+HST. Red: Retrieval using the joint-fit TESS data; Blue: Retrieval using the weighted averaged TESS data.}
    \label{fig:post_tess}
\end{figure*}

\end{document}